%% file: main.tex
\documentclass[fleqn,usenatbib]{mnras}
\usepackage{newtxtext,newtxmath}
\usepackage{fix-cm}
\usepackage{subcaption}
\usepackage[T1]{fontenc}
\DeclareRobustCommand{\VAN}[3]{#2}
\let\VANthebibliography\thebibliography
\def\thebibliography{\DeclareRobustCommand{\VAN}[3]{##3}\VANthebibliography}
\usepackage[utf8]{inputenc}
\usepackage{geometry}
\usepackage{appendix}
\usepackage{comment}
\usepackage{color}
\usepackage{graphicx}
\usepackage{orcidlink}
\usepackage{amsfonts,latexsym,epsfig,amsmath,mathrsfs}

\def\mpcoh{\,h^{-1}{\rm Mpc}}
\def\hompc{\,h\,{\rm Mpc}^{-1}}

\newcommand{\SA}[1]{\textcolor{red}{#1}}

\def\citejap#1{\citeauthor{#1}\ \citeyear{#1}}

\title[CMB lensing tomography with clustering redshifts]{CMB lensing tomography with clustering estimation of lens redshift distributions}

\author[Shun Arai et al.]{
Shun Arai\,$^{1}$\thanks{E-mail: shunarai@kmi.nagoya-u.ac.jp}\orcidlink{0000-0002-2527-3705}, John. A.  Peacock\,$^{2}$\orcidlink{0000-0002-1168-8299},
Hironao Miyatake\,$^{1,3,4}$\orcidlink{0000-0001-7964-9766}, and
Atsushi J. Nishizawa$^{5,1,3}$\orcidlink{0000-0002-6109-2397}
\\
$^{1}$ Kobayashi-Maskawa Institute for the Origin of Particles and the Universe (KMI), Nagoya University, Nagoya, 464-8602, Japan\\
$^{2}$
Institute for Astronomy, University of Edinburgh, Royal Observatory Edinburgh, Blackford Hill, Edinburgh EH9 3HJ, UK\\
$^{3}$Institute for Advanced Research, Nagoya University, Nagoya 464-8601, Japan
\\
$^{4}$Kavli Institute for the Physics and Mathematics of the Universe (WPI)\\
$^{5}$Gifu Shotoku Gakuen University, Gifu 501-6194, Japan\\
}
\date{Accepted XXX. Received YYY; in original form ZZZ}

\pubyear{2024}

\begin{document}
\label{firstpage}
\pagerange{\pageref{firstpage}--\pageref{lastpage}}
\maketitle

\begin{abstract}
We develop a clustering-based redshift estimation approach for CMB lensing tomography, focusing on the kernel function of the lensing galaxies. Within a linear galaxy bias framework, we derive estimators for this kernel from two-point cross-correlations between lens mass and reference samples. The reconstructed kernel then enables a theoretical prediction for the angular cross-power spectrum \(C_{g\kappa}\) between CMB lensing convergence and lens galaxies. As a proof of concept, we measure \(C_{g\kappa}\) by correlating the \emph{Planck} PR4 convergence map with NVSS+SUMSS radio galaxies (\(0\lesssim z\lesssim 3\)). We estimate the radio-galaxy kernel by collectively cross-correlating their distribution with spectroscopic and photometric surveys (2MPZ, LOWZ-CMASS, eBOSS DR16 LRGs, and Gaia-unWISE QSOs). From the measured \(C_{g\kappa}\), we obtain \(\sigma_8 = 0.86^{+0.12}_{-0.09}\) when the density parameter is set to the {\it Planck} value of $\Omega_m = 0.315$; this is in good agreement with the \emph{Planck} normalisation of $\sigma_8 = 0.812$.
\end{abstract}


\section{Introduction}
\label{sec:intro}

Statistical cross-correlation analysis of the cosmic microwave background (CMB) and the large-scale structure (LSS) of the Universe has been developed to reveal the growth history of the Universe over the past 13.8 billion years, testing the concordance cosmology i.e. the flat $\Lambda$CDM model based on General Relativity. In the case of weak gravitational lensing of the CMB, the CMB field is distorted into a non-Gaussian state from the Gaussian primary CMB field due to a relativistic gravitational deflection sourced by the LSS. Thus the non-Gaussian component of the CMB indicates the impact of the foreground LSS, and various techniques to reconstruct the CMB lensing convergence field from the non-Gaussian signal have been developed: e.g. \cite{2002ApJ...574..566H, Lewis:2006fu,  Maniyar:2021msb}. At present, several CMB lensing convergence maps
have become available for practical usage, enabling a range of high-precision statistical analyses: {\it Planck} PR4 \citep{Carron:2022eyg};  Atacama Cosmology Telescope (ACT) DR6 \citep{2024ApJ...962..113M}; and South Pole Telescope (SPT) \citep{2017ApJ...849..124O}.

As is well established from \cite{Lewis:2006fu}, 
the CMB lensing convergence field integrates the line-of-sight overdensity of matter inhomogeneities weighted with the dimensionless lensing kernel function, projected on the celestial sphere. In principle,  the LSS field at a localised redshift can be used to extract the CMB convergence at that redshift: the cross-correlation of the CMB convergence and the LSS  yields a CMB lensing tomography of the growth of LSS. The CMB lensing kernel function has a peak at $z \approx 2$ but is broadly distributed from $z \approx 0.1$ and its tail decreases only gradually at the higher redshift end, so there are various opportunities to construct CMB lensing tomography over a wide range of redshifts. In addition, the CMB lensing convergence field is an all-sky quantity. Thus the LSS fields with larger sky coverage can provide higher signal-to-noise for CMB lensing tomography, especially for the linear growth of the LSS. 

Measurements of CMB lensing in statistics of angular cross-correlations started with radio galaxy samples from the NRAO VLA Sky Survey (NVSS) and the CMB map provided by the Wilkinson Microwave Anisotropy Probe (WMAP) in \cite{Smith:2007rg}. These measurements of cross-correlations between LSS and the CMB lensing convergence were confirmed by various authors \cite{PhysRevLett.107.021301,2012ApJ...756..142V,Holder:2013hqu}. Regarding CMB lensing tomography, all-sky tomography up to $z \lesssim 0.6$ was completed by \cite{Peacock:2018xlz} and at $z \lesssim 0.8$ by \cite{2021MNRAS.507..510H}, extended to higher redshift slices with unWISE quasar samples \citep{2021JCAP...12..028K} and with Gaia-unWISE QSOs \citep{2023JCAP...11..043A, 2024JCAP...06..012P}. The deepest measurement of CMB lensing was obtained by \cite{2022PhRvL.129f1301M}, using Lyman-Break Galaxies; see \cite{Wilson:2019brt} for more detail on its methodology. 

The central task of attaining CMB lensing tomography is to determine the statistical properties of the galaxies that trace the LSS, namely the line-of-sight distribution and the bias of galaxy samples. For this purpose, it is appropriate to make use of galaxy samples with known redshifts as well as covering a larger portion of the sky. As is often the case, however, the existing galaxy catalogues have limitations i.e. known redshifts but small sky coverage or large sky coverage while redshifts are unknown. One method for overcoming this shortage of galaxy catalogues with known redshifts is the estimation of the redshift distribution and bias of galaxy samples via the so-called `clustering redshift' method: using measured cross-correlation functions with samples that have known spectroscopic or photometric redshifts \citep{Newman:2008mb,Rahman:2014lfa}.

In this paper, we employ a clustering redshift analysis to derive the kernel function of lensing galaxies via a data-driven method, aiming to proceed with CMB lensing tomography. We choose radio galaxies as lensing galaxies as they are expected to be a major source of CMB lensing. Radio galaxies are indeed composed of active galactic nuclei at the centre of massive galaxies which generate an enormous outflow of energy that generates bright radio emission; typically $L_{\rm 500\, MHz}(\rm rest)>10^{27}\,{\rm W\  Hz^{-1}}$ \citep{Miley:2008ti} in luminosity. Currently, tens of millions of sources are found in existing radio surveys, e.g., NVSS \citep{Condon:1998iy}; the Sydney University Molonglo Sky Survey (SUMSS) \citep{Mauch:2003zh}; and the VLA FIRST survey \citep{Becker:1995ei}. The radio catalogues with the widest coverage have surveyed more than 80\% of the whole sky.
But spectroscopy of radio galaxies in the NVSS is generally lacking, being confined to small sub-samples such as the 150 galaxies in CENSORS \citep{Best:2003sb,Brookes:2008fw}. For all surveys, the local part of the redshift distribution can be deduced from spectroscopic galaxies in the SDSS \citep{1997ApJ...475..479W, 2015ApJ...801...26H}.), but redshifts for fainter and more distant galaxies are lacking. At this point, the clustering redshift analyses are crucial in order to estimate the full form of the radio-galaxy redshift distributions.

The redshift distribution for radio galaxies is estimated using the following set of spectroscopic and photometric samples: the 2MASS Photometric Redshift catalogue
(2MPZ: \cite{2014ApJS..210....9B}); LOWZ-CMASS plus eBOSS DR16 galaxies \citep{2016AJ....151...44D}; and Gaia-unWISE photometric QSOs \citep{2024ApJ...964...69S},  applying a clustering redshift analysis over $0 \lesssim z \lesssim 3$ in order to trace the higher-redshift tail of the radio kernel. 
Concretely applying the methodology to an all-sky combined radio catalogue of 1.4 GHz NVSS-SUMSS radio samples covering 80$\%$ of the sky, we divide the sample into a foreground part that is reconstructed with the spectroscopic and photometric samples and a residual part that comes purely from the highest redshifts beyond the range of the calibration samples.  We then measure the angular cross-correlation of CMB lensing convergence in {\it Planck} PR4 map and the NVSS-SUMSS radio map in parallel with the clustering redshift analysis, constraining $\sigma_8$, the amplitude of cosmic density fluctuations at the present.

The remainder of this paper is constructed as follows. In Sec.~\ref{sec:theory}, we develop the theory of CMB lensing and introduce the statistics with auto- and cross-correlations in angular scales for galaxy distributions and the CMB lensing. In Sec.~\ref{sec:CMB-lensing_tomography_with_rec_kernel}, we explain the clustering redshift analysis and derive the estimators of the lensing galaxy kernel function in either configuration space or harmonic space 
and we explain CMB lensing tomography in harmonic space. In Sec.~\ref{sec:dataset}, we provide the data we use in the analysis. In Sec.~\ref{sec:CMBlensing_tomography_rec_NVSS-SUMSS_kernel}, we demonstrate the clustering redshift analysis in deriving the lensing galaxy kernel function and proceed with a CMB lensing tomography analysis using the derived kernel function. In Sec.~\ref{sec:conclusion}, we sum up our paper and give some discussion on the prospects for future studies of this type. In the Appendix, we supply detailed analyses of intrinsic parameters that might affect the main results. Throughout the paper, we set the fiducial cosmology by the flat $\Lambda$CDM model, giving the parameters obtained by \textit{Planck} 2018 as $(h, \Omega_{\rm b}h^2,\Omega_ch^2, {\rm ln}(10^{10}A_s), n_s) = (0.67, 0.0023, 0.12, 3.045, 0.964)$. Note that the amplitude of the power spectrum can be given by $\sigma_{8,{\rm fid}}=0.812$ as an alternative to $A_s$.

\input{CMBlens_with_RadioGals}

%
\section{Conclusions}\label{sec:conclusion}
We have discussed the method of CMB lensing tomography empowered by clustering redshift estimation applied to lensing galaxies, estimating the amplitude of the cosmological density fluctuation at present i.e. $\sigma_8$. We demonstrated our method with the existing dataset of the {\it Planck} PR4 convergence map and the radio galaxies of NVSS-SUMSS all-sky samples representing the foreground lenses. We derived the kernel function of the radio galaxies with the help of the measured correlation functions with reference samples that have known redshifts, i.e. LOWZ-CMASS, eBOSS LRG, 2MPZ, and Gaia-unWISE QSOs, applying the clustering-based redshift estimation method. Based on the reconstructed kernel function of the NVSS-SUMSS radio galaxies, we constrained $\sigma_8$ using two different models of CMB lensing kernel derived in this way. We confirmed that the $\sigma_8$ values derived from the two models are consistent with the fiducial  \textit{Planck} 2018 TT-TE-EE + Low-E value i.e. $\sigma_8 = 0.812$. We found that the model with the better prescription of the radio distribution gave the tighter constraint: $\sigma_8 = 0.86^{+0.12}_{-0.09}$.

We estimated the cumulant of the CMB lensing cross-power spectrum, $C_{g\kappa}(\ell,z_{\rm sep.})/C_{g\kappa}(\ell)$, in order to clarify the issue of a possible residual contribution from the radio galaxies at $z > z_{\rm sep.}$ given the fitting function of the reconstructed radio kernel. We found that the cumulant does not reach unity up to $z \approx 4$. This indicates that there may be a significant correlation amplitude with radio samples which comes from such high redshifts, leaving a potential measurement of such high-redshift tomographic angular cross-power spectrum in future CMB experiments and radio surveys. At present, however, the dataset of {\it Planck} PR4 and the NVSS-SUMSS radio galaxies could not show any residual signals with a significant signal-to-noise ratio at $z \approx 2$ within the measurement error of $C_{g\kappa}(\ell)$, as shown in Fig.~\ref{fig:separable_redshifts_given_noise_from_cgk_PR4_NVSS-SUMSS}. 

We have estimated $\sigma_8$ with our method and found that the value is consistent with that deduced by {\it Planck} 2018, as well as other existing measurements: cf. \cite{2021JCAP...12..028K,2023JCAP...11..043A, 2024JCAP...06..012P}. 
To this end, we conclude that there seems no significant evidence against the fiducial $\Lambda$CDM model. Meanwhile, our approach in this paper will be more desirable in probing the Universe at high redshifts for the following reasons. One is that the highest redshifts will not be well surveyed until deeper galaxy catalogues and more precise CMB lensing maps have been obtained.
The other is the mismatch of the measurements of the Hubble-Lema\^itre constant $H_0$ \citep{Verde:2019ivm} and the tension over the amplitude of density contrast $\sigma_8$ \citep{2023PhRvD.108l3517M} are not reconciled at present. In addition, non-standard theories of gravity alternatively explain the cosmic acceleration especially at a late time, speculating whether or not gravity follows the law we know both at low and high redshifts  (see comprehensive reviews for this field: \citejap{2016RPPh...79d6902K};  \citejap{2019ARA&A..57..335F,2019arXiv190509687I};  \citejap{2023PTEP.2023g2E01A}). Moreover, there is an advantage that a data-driven reconstruction of cross-correlation signals is robust to alternative cosmological models without the need to assume a particular  theory of gravity. To this end, we will proceed with our methodology preparing forthcoming Simons Observatory \citep{2019JCAP...02..056A}, CMB Stage-IV experiments \citep{2022arXiv220308024A} and Square Kilometre Array Observatory \citep{2015aska.confE.174B} in the same timeline of the Vera. C. Rubin Observatory \citep{2019ApJ...873..111I}, and the Nancy Grace Roman Space Telescope \citep{2015arXiv150303757S}.

\section*{Acknowledgements}\label{sec:acknowkedgements}
SA and JAP thank Antony Lewis for sharing the {\it Planck} PR4 convergence map.
SA thanks Toshiya Namikawa and Takahiro Nishimichi for giving helpful comments. SA and HM thank David Alonso and Giulia Piccirilli for the fruitful discussion at the CMB$\times$LSS workshop at the Yukawa Institute of Theoretical Physics in 2023. SA was supported by the Japan Society for the Promotion of Science
(JSPS) Grants-in-Aid for Scientific Research (KAKENHI) Grant No. JP24K17045. 
HM was supported by JSPS KAKENHI Grand Numbers JP20H01932,
JP23H00108, and 22K21349, and Tokai Pathways to Global
Excellence (T-GEx), part of MEXT Strategic Professional
Development Program for Young Researchers. AJN was supported by JP21K03625, JP21H05454, JP22K21349 and JP23H00108.  For open access, the
authour has applied a Creative Commons Attribution (CC BY) licence to any Author Accepted Manuscript version arising from this submission.

\section*{Data Availability Statement}\label{sec:data_availability_statement}
The data underlying this article are available in third-party data resources linked, and the authors can provide any more detail of the analyses with upon request. 

\bibliographystyle{mnras}
\bibliography{bibliography.bib}
\appendix
\input{App}
\bsp	
\label{lastpage}
\end{document}

%% file: CMBlens_with_RadioGals.tex
\section{Theory}\label{sec:theory}
We now summarise the theoretical description of the statistical quantities involved in CMB lensing and galaxy distributions. Following the formulation in \cite{Lewis:2006fu}, we describe CMB lensing 
as the solution of the geodesic motion of the CMB photons given the inhomogeneous gravitational potential and the spatial curvature in a time-evolving universe. We assume a homogeneous and isotropic universe described by the flat $\Lambda$CDM model as a background:
\begin{equation}\label{eq:bg_cosmology_flat_LCDM}
    H(z) = H_0\sqrt{\Omega_m (1+z)^3 + 1-\Omega_m}\,.
\end{equation}
Here $H(z)$ is the Hubble parameter at redshift $z$, and $\Omega_m$ is the matter-density fraction in the universe at present. Note that we ignore the radiation component $\Omega_r$, focusing on the epoch well after the last scattering of CMB. The homogeneity and isotropy at the largest scales are consistent with the observational data \citep{1990ApJ...354L..37M}. The flatness of the Universe is assumed following the measurement of the position of the first acoustic peak of CMB temperature anisotropy \citep{Spergel:2003cb,Ade:2013sjv, Planck:2018vyg}. We define the comoving distance as $\chi(z) = \int^z_0{dz^\prime/H(z^\prime)}$.
$c$ denotes the speed of light. Hereafter we omit the argument $z$ from $H$ and $\chi$ for concise notation except where we need to discuss this dependence explicitly.

\subsection{Field quantities}\label{ssec:field_quantities}
We then describe the growth of LSS on a background using perturbation theory relative to the background solution. The matter-density contrast $\delta(\chi,{\bf \hat{n}}) \equiv \rho_{\rm m}(\chi,{\bf \hat{n}})/\bar{\rho}_{\rm m}(\chi) - 1$, where ${\bf \hat{n}}$ is the angular position on the sky, is the key quantity characterising the LSS. We assume that the density contrast is linear
i.e. $\delta = D_+(z) \delta_{\rm ini.}$ with the linear growth factor $D_+$. 
We describe a galaxy field based on the scale-independent linear bias model.
We define the three-dimensional galaxy number fluctuation for a given galaxy sample as $\delta_g (\chi,{\bf \hat{n}}) \equiv n_g(\chi,{\bf \hat{n}})/{\bar{n}_g(\chi)}-1$, where $n_g$ and $\bar{n}_g$ denote the number density and the mean number density of galaxies, respectively. Under the linear bias approximation, we can write the relation between the galaxy number fluctuation and the matter-density contrast as
\begin{equation}\label{eq:delta_G_3D}
    \delta_g(\chi,{\bf \hat{n}}) = b_g (z) \delta(\chi,{\bf \hat{n}})\,,
\end{equation}
where $b_g(z)$ is the scale-independent linear bias of the galaxy. The two-dimensional galaxy field is defined by the following integration:
\begin{equation}\label{eq:delta_G_2D}
    \delta_g({\bf \hat{n}}) = \int^{z_{\rm max.}}_0{{\rm d}
z\, K_g (z
)\delta(\chi,{\bf \hat{n}})}\,,
\end{equation}
with the galaxy kernel function along the line of sigh:
\begin{equation}\label{eq:kernel_G}
    K_g(z) = \frac{b_g(z)}{\bar{n}_g(z)}\frac{{\rm d}\bar{n}_g}{{\rm d}z}(z)\,.
\end{equation}
Here we change the argument of the functions from $\chi$ to $z$.\footnote{Throughout this paper, we define a dimensionless kernel function $K_A$, which differs by $H/c$ factor from $W_A$ of the ordinary definition of the kernel function in the literature. }. The upper limit of the integral is determined by the maximum redshift of the galaxy included in the sample i.e. $z_{\rm max.}$.

The linear bias model 
well approximates the matter fluctuation on large scales, but it is invalid for $k > 0.1\hompc$ where scale dependent bias arises due to the nonlinear evolution of the structures.
Hence it is necessary to examine whether the scale-independent linear bias model is applicable; we investigate this issue for our sample in App.\ref{app:validation_of_linear_bias_model}, 
showing that the derived bias factor is consistent with scale independence as long as we properly model the nonlinear dark matter overdensity via HALOFIT \citep{2003MNRAS.341.1311S, 2012ApJ...761..152T}.

The CMB lensing convergence field $\kappa({\bf \hat{n}})$ is given as
\begin{equation}\label{eq:kappa_2D}
    \kappa({\bf \hat{n}}) = \int^{z_{\rm *}}_0{{\rm d}z K_\kappa (z) \delta(\chi,{\bf \hat{n}})}\,,
\end{equation}
where $z_*$ is the redshift at the last scattering surface of CMB. Note that $\kappa({\bf \hat{n}})$ is by definition a two-dimensional field the gravitational lensing performs a projection from the last scattering surface to the celestial sphere.
Given the Poisson equation and taking a small-angle approximation, the CMB lensing kernel is derived as
\begin{equation}\label{eq:kernel_kappa}
    K_\kappa (z)\equiv \frac{3\Omega_m H^2_0}{2c H}\frac{(1+z)(\chi_* - \chi)\chi}{\chi_*}\,,
\end{equation}
where $\chi_* \equiv \chi(z_*)$ is the comoving distance to the last scattering surface. Note that the comoving distance appears as we assume the global spatial curvature to be zero.

\subsection{Statistical angular correlations}\label{ssec:stats_corrs}
Summary statistics help extract cosmological information from the LSS. The angular auto-correlation of galaxies provides the underlying matter clustering, although galaxies are biased tracers of the underlying dark matter distribution.
The angular cross-correlation of galaxies at a certain redshift and a convergence field of CMB lensing enable us to extract the dark matter distribution at the redshift of the galaxy sample. Note that the dark matter distribution measured from the cross-correlation is still around the biased tracer.
We briefly summarise the definition of these angular correlations in the following subsections.

\subsubsection{Configuration-space correlation function}\label{sssec:config_sp_corr}
A projected correlation function is defined as
\begin{equation}\label{eq:wAB_def}
    w_{AB}(\theta) \equiv \left< \delta_A({\bf \hat{n}}) \delta_B({\bf \hat{n}}')\right>_{\rm ens.}\,,
\end{equation}
where $\delta_A({\bf \hat{n}})\equiv\int{{\rm d}z K_A(z)\delta(\chi,{\bf \hat{n}})}$ and the subscripts {\it A} and {\it B} denote the types of fields, and $\left<\cdot\right>_{\rm ens.}$ denotes the ensemble average 
over hypothetical universes. 
Under the assumption that $\theta$ is small and the sky is flat, we can apply the Limber approximation \citep{1953ApJ...117..134L} and obtain
\begin{equation}\label{eq:wAB_lof_proj}
    w_{AB}
    (\theta) = \int^{\infty}_0{{\rm d}z\,}{K_A(z)K_B(z)w(\chi\theta)}\,.
\end{equation}
Here $w(\chi\theta)$ is the projected two-point correlation function of matter,
\begin{equation}\label{eq:w_proj_def}
    w(z, \theta; \pi_{\rm max}) \equiv \frac{2H(z)}{c}\int^{\pi_{\rm max}}_0{\rm d}\pi\,\xi\left(\sqrt{(\chi\theta)^2 + \pi^2}\right)\,.
\end{equation}
$\pi_{\rm max}$ denotes the maximum projection length along the line of sight. $\xi (r)$ is the three-dimensional two-point correlation function of matter-density contrast, $\xi(r) \equiv \left< \delta_(\chi,{\bf \hat{n}}) \delta(\chi',{\bf \hat{n}}')\right>_{\rm ens.}  = \int{{\rm d}kk^2P_m(k,z)j_0(kr)}/2\pi^2$ with the matter-power spectrum $P_m(k,z)$ and the zeroth spherical Bessel function $j_0(x)$. Recalling that we assume the matter-power spectrum keeps translation and rotational invariance, the two-point correlation function only depends on the comoving separation, $r \equiv |\chi{\bf \hat{n}} - \chi'{\bf \hat{n}}'|$.
In practice $\pi_{\rm max}$ is appropriately chosen so that the systematics from the peculiar velocities of galaxies i.e. the effect of redshift-space distortions (RSD) is not effective, namely confirming that the RSD correction from the Kaiser RSD factor is negligible; cf. \cite{2010MNRAS.407..520N}.  
We will describe how specifically we choose the values for $\pi_{\rm max.}$ given the dataset in Table.~\ref{tab:def_of_zslices} in Sec.~\ref{sec:dataset}.

\subsubsection{Harmonic-space correlation function}\label{sssec:harmo_sp_corr}
We employ the harmonic-space angular power spectrum following the convention of CMB measurements e.g. \cite{Planck:2018vyg}, which also has become common in galaxy surveys \citep{2019PASJ...71...43H, 2021MNRAS.505.5714A, 2022A&A...665A..56L, 2024MNRAS.528.2112S, 2024ApJ...966..157F}. The projected field variable is expanded as $\delta_A({\bf \hat{n}}) = \sum^{\infty}_{\ell = 0}\sum^{\ell}_{m=-\ell}a_{A,lm}Y_{\ell m }({\bf \hat{n}})$ in terms of the spherical harmonics $Y_{\ell m }({\bf \hat{n}})$. Then we define the angular power spectrum between the $A$ and $B$ fields as
\begin{equation}\label{eq:clAB_def}
    C_{AB}(\ell) \equiv (2\ell + 1)^{-1}\sum^{\ell}_{m=-\ell}\left<a_{A,\ell m} a_{B,\ell m} \right>_{\rm ens.}\,.
\end{equation}
As long as the Limber approximation is valid \citep{1953ApJ...117..134L,1992ApJ...388..272K} we obtain $C_{AB}$ as
\begin{equation}\label{eq:clAB_limber}
    C_{AB}(\ell) = \int^\infty_0{{\rm d}z} \frac{K_A(z)K_B(z)}{\chi^2c/H} P_m\left(\frac{\ell+1/2}{\chi},\,z\right)\,,
\end{equation}
which computes the integral of the matter-power spectrum  weighted by kernel functions along the line-of-sight. 

\subsection{Amplitude of cosmological density fluctuations}\label{ssec:def_sigma8}
The amplitude of cosmological matter-density fluctuations is the key quantity characterising how the LSS grows in cosmic history. We define the amplitude parameter $\sigma$ through
\begin{align}\label{eq:def_amp_delta}
    &\sigma^2(R,z) \equiv \int^\infty_0{\frac{dk k^2}{2\pi^2} P^{\rm lin.}_m(k,z)W^2(kR)}\,,\\ \nonumber
    & W(kR) \equiv \frac{3j_1(kR)}{kR}\,,
\end{align}
where $j_1(x)$ is the first spherical Bessel function and $P^{\rm lin.}_m(k,z)$ is the linear matter-power spectrum. Note that $W(x)$ is the Fourier transform of the top-hat window function in configuration space, so that $\sigma^2$ denotes a smoothed fluctuation at a finite radius $R$. We introduce $\sigma_8 \equiv \sigma(R=8[\!\mpcoh],z=0)$ representing the fractional linear rms matter fluctuation.
Note that the statistical two-point functions $w_{AB}$ in Eq. ~\eqref{eq:wAB_lof_proj} and $C_{AB}$ in Eq.~\eqref{eq:clAB_limber} are proportional to $\sigma^2_8$.

\section{CMB lensing tomography with reconstructed kernel function}\label{sec:CMB-lensing_tomography_with_rec_kernel}
The conventional analysis for CMB lensing tomography requires a knowledge of 
the redshift distribution of lens galaxies and in addition one has to determine the value of the linear bias of the lens galaxies in order to complete the theoretical interpretation of galaxy-CMB lensing measurements. In this methodology, the auto-correlation function of the lens galaxies is the key quantity for disentangling the linear bias from $\sigma_8$. However, some lens samples lack a redshift estimation for each galaxy, while others only have noisy measurements of auto-correlation functions due to systematics. These are issues that we must confront when using radio galaxies as a lens sample for CMB lensing tomography. 

To avoid these obstructions and enhance the utility of radio samples to constrain $\sigma_8$, we use the so-called clustering redshift method. This approach does not provide a redshift estimation for each galaxy but enables us to quantify the kernel function of the lens galaxies given a certain range of redshifts, namely the multiplication of the linear bias and the redshift distribution of our lens sample, i.e., $b_g{\rm d}\bar{n}_g/{\rm d}z(z)$ in Eq.~\eqref{eq:kernel_G}. Regarding the advantage, clustering-based analysis directly provides the kernel function from measured correlation functions and thus one does not have to assume a model of the kernel or even have to separate $b_g$ and ${\rm d}\bar{n}_g/{\rm d}z$ in computing a theoretical prediction of CMB lensing tomography with radio galaxies. We will describe in detail in
Sec.~\ref{ssec:CMB lensing_tomography_with_rec_Ku} how this approach can be advantageous for predicting the angular cross-power spectrum between lens galaxies and the CMB lensing convergence.

\subsection{Clustering-based redshift estimation}\label{ssec:clustering_z}
The clustering redshift method was originally established by \cite{Newman:2008mb}, and later refined by  \cite{Rahman:2014lfa} to determine the lensing kernel function. The clustering-based redshift estimation relies on the fact 
that two different galaxy samples located in the same redshift are associated with the same underlying matter fluctuations.
As a result, the redshift distribution of a galaxy sample without known redshifts can be estimated through the cross-correlation between the galaxy sample with unknown redshifts and galaxy samples with well-known redshifts.

We formulate the reconstruction of the kernel function which is expressed using measured correlation functions.
We now consider the situation that there are two different types of galaxy catalogues, where one has well-known redshifts and the other has unknown redshifts, and they have an intersection in volume. We denote the former as a reference sample and the latter as an unknown-redshift sample, in short, $r$ and $u$, respectively. 

\subsubsection{Narrow redshift slices}\label{sssec:narrow_z_slices}
We first consider the case where the intersections are well-localised in redshifts. We divide the sample into the redshift bins so that each bin consists of an equal number of galaxies i.e. $N$ out of the total galaxy number $N_r$. Then we make a histogram of the reference galaxy as
\begin{equation}\label{eq:dndr_ref_narrow}
    \frac{1}{n_r(z)}\frac{{\rm d}n_r}{{\rm d}z}(z) = \frac{1}{N_b}\sum^{N_b}_{i=1}\frac{1}{\Delta z_i}\Pi\left(\frac{z - z_i}{\Delta z_i}\right) + {\cal O}(\Delta z^2)\,.
\end{equation}
Here $z_i$ denotes the centre of the $i$-th bin and the width of redshift bin ${\Delta z}_i$\ for $i=1,2,\cdots,N_b$ with $N_b \equiv [N_r/N]$. $\Pi(x)$ is the rectangular function defined as
\begin{align}\label{eq:rectangular_func}
    \Pi(x) = 
    {\begin{cases}
    1 \ &(-1/2 < x < 1/2) \\
    1/2 \ &(x = \pm 1/2)\\
    0 \ &(x < -1/2, x > 1/2)
    \end{cases}}
\end{align}
Note that the correction starts from the second-order in $\Delta z$ due to the symmetry of the rectangular function. The correction at ${\cal O}(\Delta z^2)$ becomes more negligible as the bin widths get narrower. 
The equality satisfied up to the order ${\cal O}(\Delta z)$ is hereafter denoted as $\approx$. The bin widths differ in different bin numbers. The order of magnitude of the bin widths is expressed by $N$, $N_r$, and a certain redshift span of the sample $z_{\rm range}$ as
\begin{equation}\label{eq:delta_z_i}
   \Delta z_i  = {\cal O}\left(\frac{N}{N_r}z_{\rm range}\right)\,,
\end{equation}
provided that the distribution does not decay sharply in the range of $z_{\rm range}$. Substituting the approximated distribution Eq.~\eqref{eq:dndr_ref_narrow} into the kernel function of Eq.~\eqref{eq:delta_G_3D} and calculating the correlation function, Eq.~\eqref{eq:wAB_lof_proj}, we obtain the formulae of the correlation functions for $i$-th redshift bin as
\begin{align}
 w_{rr} (z_i,\theta) &\approx \frac{b^2_r (z_i)}{\Delta z_i}w(z_i,\theta,\pi_{{\rm max},i})\,,\label{eq:wrr_i}\\
    w_{ur} (z_i,\theta) &\approx b_r(z_i)K_u(z_i)w(z_i,\theta,\pi_{{\rm max},i})\,,\label{eq:wur_i}
\end{align}
where $\pi_{{\rm max},i} \equiv \Delta z_i c/2H(z_i)$. We analytically solve the equations for $K_u(z_i)$ by eliminating the $b_r$ term and obtain 
\begin{equation}\label{eq:kernel_u_rec_config_sp}
    K^{\rm rec.}_u (z_i) \equiv \frac{w_{ur} (z_i,\theta)/w(z_i,\theta,\pi_{{\rm max},i})}{\sqrt{w_{rr} (z_i,\theta)\Delta z_i/w(z_i,\theta,\pi_{{\rm max},i})}}\,.
\end{equation}
Note that the scale dependence vanishes as the dark matter correlation function cancels in the combinations $w_{rr}/w$ and $w_{ur}/w$, respectively. This suggests that we obtain estimators of the kernel function at the individual angular bins where the correlation functions are measured, reducing the statistical error on $K^{\rm rec.}_u$. We will consider in detail the estimation of $K^{\rm rec.}_u$ in Sec.~\ref{sssec:est_config-sp_analysis}.
The advantage of the estimator Eq.~\eqref{eq:kernel_u_rec_config_sp} is that one does not have to separate the bias and the redshift distribution for the unknown sample, directly calculating the projected correlation functions with the derived kernel. This advantage is particularly helpful for radio samples for which the bias and redshift distributions are less well established than for other spectroscopic/photometric galaxy samples.

In harmonic space, we similarly derive the relations between the observables and the matter power spectrum as

\begin{align}
 {C}_{rr} (z_i,\ell) &\approx \frac{b^2_r (z_i)}{\Delta z_i}C(z_i,\ell)\,,\label{eq:clrr_i}\\
 {C}_{ur} (z_i,\ell) &\approx b_r(z_i)K_u(z_i)C(z_i,\ell)\,,\label{eq:clru_i}
\end{align}
where $C$ denotes 
\begin{equation}\label{eq:cl_intg}
    C(z,\ell) \equiv \frac{P_m((\ell+1/2)/\chi(z),z)}{\chi^2(z)c/H(z)}\,.
\end{equation}
We analytically solve the equations for $K_u(z_i)$ and obtain 
\begin{equation}\label{eq:kernel_u_rec_harmo_sp}
    K^{\rm rec.}_u (z_i) \equiv \frac{C_{ur} (z_i,\ell)/C(z_i,\ell)}{\sqrt{C_{rr} (z_i,\ell)\Delta z_i/C(z_i,\ell)}}\,,
\end{equation}
which gives the harmonic-space complement to Eq.~\eqref{eq:kernel_u_rec_config_sp}. Siminar to Eq.~\eqref{eq:kernel_u_rec_config_sp}, the multipole dependence vanishes for $K^{\rm rec.}_u$ and thus there are estimators of the kernel function individually obtained at the multipole bands where the angular power spectrum are measured, reducing the statistical error on $K^{\rm rec.}_u$. We will discuss the estimation of $K^{\rm rec.}_u$ in Sec.~\ref{sssec:est_harmo-sp_analysis}.

\subsubsection{Broad redshift slices}\label{sssec:broad_z_slices}
We derive the approximated kernel functions from the correlation functions when an individual galaxy in the reference sample has a 
probability distribution with a significant spread. 
Formulating this case is necessary in practice for the cases where the number density of the reference in certain redshift bins is too low to trace the high signal to noise of correlation statistics, or where there are limiting uncertainties in  photometric redshifts of reference samples. 

Let us define the following quantities that characterise the peak and the width of the probability distribution of a reference sample: ${\rm d}p_r/{\rm d}z \equiv n^{-1}_r {\rm d}n_r/{\rm d}z$ as
\begin{align}
    &z_{\rm mean} \equiv E[z]\,,\label{eq:zmean_via_E}\\
    &\sigma^2_z \equiv E\left[(z-z_{\rm mean})^2\right]\,,\label{sigma2_z_via_E}
\end{align}
with the functional $E[X(z)]\equiv\int{{\rm d}z X(z)({\rm d}p_r/{\rm d}z)}$. Suppose that ${\rm d}p_r/{\rm d}z$ is well approximated by a Gaussian distribution centred on $z_{\rm mean.}$, with a standard deviation $\sigma$. In this case, the measured correlation functions are expressed via the theoretical models as
\begin{align}
    &w_{rr} = E[b^2_r({\rm d}p_r/{\rm d}z) w]\,,\label{eq:wrr_broad_z}\\
    &w_{ur} = E[b_rK_u w]\,.\label{eq:wur_broad_z}
\end{align}
In harmonic space, we obtain
\begin{align}
    &C_{rr} = E[b^2_r({\rm d}p_r/{\rm d}z)C]\,,\label{eq:clrr_broad_z}\\
    &C_{ur} = E[b_rK_u C]\,,\label{eq:clur_broad_z}
\end{align}

In what follows, let us validate the above estimators given reasonable assumptions. In comparison to the narrow slices, which correspond to the limit $\sigma_z \rightarrow 0$, one can obtain $E(X(z)) \rightarrow X(z)$ and 
${\rm d}p_r/{\rm d}z \rightarrow \delta (z - z_{\rm mean})$
, reproducing the formulae for the narrow redshift slices at the order of $\sigma^2_z$. One may relax the required sharpness of ${\rm d}p_r/{\rm d}z$ to
\begin{equation}\label{eq:cond_kernel_rec_broad_z}
    b^{-1}_r|{\rm d}b_r/{\rm d}z|, K^{-1}_{u}|{\rm d}K_{u}/{\rm d}z| \ll {\rm d}p_r/{\rm d}z\,. 
\end{equation}
as discussed in \cite{Rahman:2014lfa}. This allows us to pick the kernel function out of the integral. We obtain the following approximate relations for the correlation functions in configuration space as
\begin{align}
    &w_{rr}(z_{\rm mean},\theta) \approx b^2_r(z_{\rm mean})E[w({\rm d}p_r/{\rm d}z)]\,,\label{eq:wrr_ito_bmean}\\
    &w_{ur}(z_{\rm mean},\theta) \approx b_r(z_{\rm mean})K_u(z_{\rm mean})E[w]\,,\label{eq:wur_ito_bKmean}
\end{align}
and in harmonic space as
\begin{align}
 &{C}_{rr} (z_{\rm mean},\ell) \approx b^2_r (z_{\rm mean})E\left[C(z,\ell)({\rm d}p_r/{\rm d}z)\right]\,,\label{eq:clrr_ito_bmean}\\
 &{C}_{ur} (z_{\rm mean},\ell) \approx b_r(z_{\rm mean})K_u(z_{\rm mean})E\left[C(z,\ell)\right]\,,\label{eq:clur_ito_bKmean}
\end{align}
introducing the mean linear bias $b(z_{\rm mean})$ and the kernel function $K_u(z_{\rm mean})$ at $z = z_{\rm mean}$. Therefore, we obtain the estimator of the kernel function as
\begin{equation}
\label{eq:kernel_u_rec_broad_z_config_sp}
    K^{\rm rec.}_{u}(z_{\rm mean},\sigma_z) \equiv \frac{w_{ur}/E[w]}{\sqrt{w_{rr}/E[w({\rm d}p_r/{\rm d}z)]}}\,,
\end{equation}
in configuration space, and
\begin{equation}\label{eq:kernel_u_rec_broad_z_harmo_sp}
    K^{\rm rec.}_{u}(z_{\rm mean},\sigma_z) \equiv \frac{C_{ur}/E[C]}{\sqrt{C_{rr}/E[C({\rm d}p_r/{\rm d}z)]}}\,,
\end{equation}
in harmonic space, respectively.
We obtain $K^{\rm rec.}_u(z_{\rm mean}) = K_u(z_{\rm mean})$ substituting Eqs.~\eqref{eq:wrr_ito_bmean} and ~\eqref{eq:wur_ito_bKmean} into Eq.~\eqref{eq:kernel_u_rec_broad_z_config_sp}, or substituting Eqs.~\eqref{eq:clrr_ito_bmean} and ~\eqref{eq:clur_ito_bKmean} into Eq.~\eqref{eq:kernel_u_rec_broad_z_harmo_sp}, as long as the condition Eq.~\eqref{eq:cond_kernel_rec_broad_z} holds.
The scale dependence of the matter correlation function is cancelled out as long as the proper model for the nonlinear matter power spectrum is assumed.

We remark on the comparison with the original clustering redshift estimation process discussed in \cite{Rahman:2014lfa}. We separately derive the narrow and broad redshift distributions.  In practice applying the estimation of the kernels, we will assign either narrow or broad redshift slices concerning the distribution of the reference samples. In addition, we find that the derived kernel functions in Eqs.~\eqref{eq:kernel_u_rec_config_sp}, ~\eqref{eq:kernel_u_rec_harmo_sp}, ~\eqref{eq:kernel_u_rec_broad_z_config_sp}, and ~\eqref{eq:kernel_u_rec_broad_z_harmo_sp} are proportional to $\sigma_{8,{\rm est.}}/\sigma_{8,{\rm fid.}}$, where $\sigma_{8,{\rm est.}}$ is the true value of the amplitude and 
$\sigma_{8,{\rm fid.}}$ is the amplitude given the fiducial cosmology. These scaling arise from inserting $w_{rr}, w_{ur}\propto \smash{\sigma^2_{8,{\rm est.}}}$ and $w \propto \smash{\sigma^2_{8,{\rm fid.}}}$ into Eqs.~\eqref{eq:kernel_u_rec_config_sp} and  ~\eqref{eq:kernel_u_rec_broad_z_config_sp} and inserting $C_{rr}, C_{ur}\propto \smash{\sigma^2_{8,{\rm est.}}}$ and $C \propto \sigma^2_{8,{\rm fid.}}$ into  Eqs.~\eqref{eq:kernel_u_rec_harmo_sp} and  ~\eqref{eq:kernel_u_rec_broad_z_harmo_sp}. This reflects the fact that one cannot know $\sigma_{8,{\rm est.}}$ a priori. 
We will carefully treat this proportionality in Sec.~\ref{ssec:CMB lensing_tomography_with_rec_Ku} and Sec.~\ref{sec:CMBlensing_tomography_rec_NVSS-SUMSS_kernel} in which $\sigma_{8,{\rm est.}}$ can be constrained in combination with the derived kernel function for the radio sample and the measured angular cross correlation between the radio sample and \textit{Planck} CMB lensing convergence.

\subsection{Estimator of CMB lensing tomography with reconstructed kernel function}\label{ssec:CMB lensing_tomography_with_rec_Ku}

We theoretically derive the angular cross-power spectrum between a lens sample and the CMB lensing convergence, given a kernel function for a lens sample that is reconstructed by the clustering-redshift analysis. We denote the lens sample with the subscript $u$ following the previous subsection. The angular cross-power spectrum between the lens sample and the CMB lensing convergence is given as
\begin{equation}\label{eq:clukappa_limber}
    C_{u\kappa}(\ell) = \int^{z_*}_0{{\rm d}z} \frac{K_u(z)K_{\kappa}(z)}{\chi^2c/H} P_m\left(\frac{\ell+1/2}{\chi},\,z\right)\,.
\end{equation}

Let us suppose that the lensing kernel function is reconstructed by reference samples at $N_{\rm rec.}$ redshift bins that are composed of both narrow slices and broad slices that satisfy the condition Eq.~\eqref{eq:cond_kernel_rec_broad_z} in redshift. Note that the number $N_{\rm rec.}$ does not include the bins missed or removed during the reconstruction. Given the reconstructed kernel, the integral on the right-hand side can be approximated by 
\begin{align}\label{eq:cl_ukappa_dec}
    C_{u \kappa}(\ell) \sim\  & \Sigma_{u\kappa}(\ell) + S_{u\kappa}(\ell)\,,\\ \nonumber
    \Sigma_{u\kappa}(\ell) \equiv& \sum^{N_{\rm rec.}}_{i=1}\frac{K^{\rm rec.}_u(z_i)K_\kappa(z_i)\delta z_i}{\chi^2(z_i)c/H(z_i)}P_m\left(\frac{\ell+1/2}{\chi(z_i)},\,z_i\right)\,,\\ \nonumber
    S_{u\kappa}(\ell) \equiv&  \int^{z_*}_{{z_{\rm max}}}{ {\rm d}z\frac{{K^{\rm res.}_u}(z)K_\kappa(z)}{\chi^2(z)c/H(z)} P_m\left(\frac{\ell+1/2}{\chi(z)},\,z\right)}\,.
\end{align}
$K^{\rm res.}_u(z)$ denotes the residual component from higher redshifts that is not reconstructed from the reference samples. Note that $\delta z_i$  need not coincide with the bin widths taken for the reconstruction, which enables one to calculate $\Sigma_{u\kappa}$ even if some bins are missed. Thus $z_i$ and $\delta z_i$ denote the representative redshift of the $i$-th bin and the width between the $i$-th and $i+1$-th bins, respectively. $z_i$ and $\delta z_i$ are equivalent to the centre of $i$-th bin and $\delta z_i = {\cal O}(\Delta z_i)$ in the case of narrow slices. On the other hand, in the case of broad slices $z_i = z_{{\rm mean},i}$ and $\delta z_i ={\cal O}(\sigma_{z,i})$. Note that $\Sigma_{u\kappa}$ becomes more accurate up to ${\cal O}(\delta z^2)$ as $\delta z_i$ gets smaller because the first-order term of  Taylor expansion of the integrand in Eq.~\eqref{eq:clukappa_limber} around the $i$-th bin centre $z = z_i$ vanishes. This indicates that one can assume
\begin{equation}\label{eq:kernel_u_of_z}
    K_u(z) \approx \sum^{N_{\rm rec.}}_{i=1}K^{\rm rec.}_u(z_i)\Pi\left(\frac{z - z_i}{\delta z_i}\right) 
    + K^{\rm res.}_u(z)\Theta(z-z_{\rm max})\,,
\end{equation}
as long as we are estimating $C_{u\kappa}(\ell)$ at the order of ${\cal O}(\delta z^2)$, provided the narrow slices are accomplished.
Here $\Theta(x)$ denotes the Heaviside function.

We find that $\Sigma_{u\kappa}$ is proportional to $\sigma_{8,{\rm est.}}\sigma_{8,{\rm fid.}}$, substituting $K^{\rm rec.}_u \propto \sigma_{8,{\rm est.}}/\sigma_{8,{\rm fid.}}$ and $P_m \propto \smash{\sigma^2_{8,{\rm fid.}}}$. On the other hand, the left-hand side $C_{u\kappa}(\ell)$ is proportional to $\smash{\sigma^2_{8,{\rm est.}}}$ as it can be measured using two-point statistics. Then we define
\begin{align}\label{eq:alpha_sigma8_estimator}
    \alpha \equiv \frac{C_{u\kappa}}{\Sigma_{u\kappa} + S_{u\kappa}}\,,
\end{align}
estimating $\sigma_{8,{\rm est.}}$. Provided $S_{u\kappa}$ is negligibly small compared to $\Sigma_{u\kappa}$, we obtain $\alpha = C_{u\kappa}/\Sigma_{u\kappa}$ and $\sigma_{8,{\rm est.}} = \alpha\sigma_{8,{\rm fid.}}$. Note that $\alpha$ now consists only of the measured two-point statistics as $S_{u\kappa}$ is ignored. In practice, however, one cannot quantify $S_{u\kappa}$ solely by observables with dependence on a dataset, introducing systematic uncertainty to measure $\sigma_{8,{\rm est.}}$. Eq.~\eqref{eq:alpha_sigma8_estimator} implies that the larger $S_{u\kappa}$ makes $\alpha$ smaller, ending up with estimating a smaller value of $\sigma_{8,{\rm est.}}$. We will assess $S_{u\kappa}$ with a specific choice of models for the kernel function given a dataset in Sec.~\ref{sec:CMBlensing_tomography_rec_NVSS-SUMSS_kernel}


\section{Dataset}\label{sec:dataset}
In this section, we provide the dataset composed of the lensing galaxies, the reference galaxy samples, and the CMB lensing convergence map, to proceed with the tomographic analysis of CMB lensing. The lens galaxies are selected from a combined radio sample consisting of NVSS \citep{Condon:1998iy} and SUMSS \citep{Mauch:2003zh}, which cover most of the sky and potentially trace the lens distribution up to $z \sim 3$ \citep{Best:2003sb,Brookes:2008fw}.
The reference galaxies are selected from the publicly available redshift samples that have a significant overlap with the lensing galaxy sample. They are spectroscopic redshift samples, LOWZ-CMASS and eBOSS DR16 LRGs, and the photometric redshift samples, 2MPZ and Gaia-unWISE quasars, spanning $0\lesssim z \lesssim 3$.  We use  \textit{Planck} PR4 \citep{Carron:2022eyg} as the latest CMB lensing convergence map, with the best understanding of calibration and noise properties. In the following subsections, we describe each dataset in more detail.

\subsection{Radio samples}
\label{ssec:radio_samples}
\subsubsection{NVSS}\label{sssec:nvss}
NVSS covers the sky north of declination $\delta > -40^\circ$ ($f_{\rm sky} \approx 0.82$) at 1.4\,GHz, containing about 1.8 million galaxies and being claimed to be 99 per
cent complete at integrated flux density $S_{1.4\,{\rm GHz}} = 3.5$\,mJy \citep{Condon:1998iy}. The resolution of images is nearly uniform with 45$''$  FWHM, which means that the majority of radio sources are unresolved. The uniformity of the source number density in declination is confirmed in the samples above 10\,mJy/beam, but fainter objects display some systematic variation that breaks the uniformity; see Fig.1 in \cite{Blake:2001bg}.
We mitigate this problem when combining the NVSS map with the SUMSS catalogue, controlling the flux cut so that the number density is close to uniform over the whole sky.

\subsubsection{SUMSS}\label{sssec:sumss}
SUMSS observations were carried out with the Molonglo Observatory Synthesis Telescope, operating at 843 MHz with a 5\,${\rm deg^2}$ field of view.
In the first data release \citep{Mauch:2003zh}, the catalogue covers approximately 3,500 ${\rm deg^2}$ of the southern sky with $\delta \leq 30^\circ$, containing about 0.1 million radio sources. The angular resolution reaches 1-2 arcsec for sources with peak brightness $A_{843} \geq 20 {\rm mJy/beam}$ and is always better than 10 arcsec. \cite{Mauch:2003zh} showed that the samples to a depth of 8 mJy at $ \delta \leq -50 ^\circ$ and 18 mJy at $\delta > -50 ^\circ $ are highly uniform. In the second data release \citep{Murphy:2007aa} the survey area is completed, extending the sky coverage to 8,100 ${\rm deg^2}$ ($f_{\rm sky} \approx 0.25$) with $\delta \leq -30 ^\circ$ and $|b| > 10 ^\circ$ i.e. the removal of the Galactic plane. The catalogue contains 210,412 radio sources to a limiting peak brightness of 6 mJy/beam at $\delta \leq -50 ^\circ$ and 10 mJy/beam at $\delta > -50 ^\circ$.

We combine the two radio catalogues so that the mean surface number density of radio objects is uniform. We compare the two catalogues in the overlapping region $-40^\circ < \delta < 30^\circ$, adjusting the lower flux limit for each catalogue so that the surface number density becomes equivalent. 
We find that the fluxes of NVSS and SUMSS are related approximately as $S_{\rm SUMSS} = S_{\rm NVSS}(0.843/1.4)^{-0.8}$, which indicates that the SUMSS sample should be chosen with a flux limit $ > 12\,{\rm mJy}$ to match NVSS (see a similar analysis consistent with ours in \citejap{Tiwari:2018hrs}).
As a result, we obtain the NVSS-SUMSS catalogue, which approximates a sample selected at 1.4\,GHz with a flux limit $> 8\,{\rm mJy}$. In what follows, we assume that the slight difference in intrinsic selection frequency causes no important difference in the redshift distribution between the NVSS and SUMSS contributions.

\subsection{Spectroscopic redshift samples}\label{ssec:specz_samples}

\subsubsection{LOWZ-CMASS sample}\label{sssec:lowz_cmass}
The LOWZ-CMASS sample consists of two distinct subsamples obtained from the SDSS Baryonic Oscillation Spectroscopic Surveys (BOSS). The LOWZ sample consists of luminous red galaxies (LRGs) up to $z \approx 0.4$ and the CMASS sample consists of massive galaxies in $0.4 < z < 0.7$.  
The LOWZ sample is defined using colour-magnitude cuts that follow the predicted track of a passively-evolving stellar population \citep{1998ApJ...500..525S}, similarly selecting the brightest and the reddest galaxies to those of SDSS-I/II Cut-I LRGs \citep{SDSS:2001wju}. The LOWZ sample extends the SDSS-I/II LRGs by selecting fainter galaxies, thereby increasing the number density.
The CMASS sample is selected by using ({\it g-r}) and ({\it r-i}) colours to isolate high redshift galaxies, which is similar to the approach of SDSS-I/II Cut-II \citep{SDSS:2001wju} and 2SLAQ LRGs \citep{2006MNRAS.372..425C}. It is a critical difference in proceeding with CMASS sample selection that the colour boundaries for the SDSS-I/II Cut-II and the 2SLAQ LRGs have been entirely removed. Alternatively, a sliding cut in colour-magnitude relations is specifically designed to collect the more massive objects as uniformly as possible as a function of redshift. 

The statistical properties of the LOWZ and CMASS samples have been well understood after carefully treating the systematics from stellar density and seeing: cf. \cite{2014MNRAS.441...24A}; for more detailed information on weights, see \cite{2016MNRAS.455.1553R} and \cite{2017MNRAS.464.1168R}. Clustering analyses of the DR12 LOWZ and CMASS samples, using two-point statistics, can be found in \cite{2016MNRAS.457.1770C} and \cite{2017MNRAS.465.1757G}, which provide the clustering signature of the LOWZ-CMASS samples at  three different bins of redshift. The random samples required for estimation of clustering signals are designed to follow the equivalent redshift distribution as the galaxy samples, and have been corrected with the same weighting as the galaxy samples. We will calculate the two-point correlation functions between the NVSS-SUMSS sample and the LOWZ-CMASS sample in configuration space, taking advantage of the fact that the masks are well characterised in both cases.
\vspace{-0.4cm}
\subsubsection{eBOSS DR16 LBGs}\label{sssec:eBOSS_DR16_LRGs}
The extended Baryon Oscillation Spectroscopic Survey \citep[eBOSS][]{2016AJ....151...44D} is the successor of BOSS in the fourth generation of the SDSS \citep{2017AJ....154...28B}. eBOSS traces Emission Line Galaxies (ELGs) and quasars (QSOs), as well as new LRGs in the redshift range $0.6 < z < 1.0$ and $z_{\rm eff}=0.698$ combined with the previous BOSS sample \citep[cf.][]{2020MNRAS.498.2354R}. The eBOSS DR16 LRGs is composed of 202,642 CMASS samples in $0.5 < z < 0.75$ and 174,816 eBOSS LRGs. The random samples consist of 50 times more objects than in the data to minimize the shot noise contribution in the estimated correlation function, with redshifts being randomly sampled from  the data. The spectroscopic information is finally matched to the remaining targets; see \cite{2021MNRAS.500..736B} for more detail. We calculate the cross-correlation between the NVSS-SUMSS sample and the LOWZ LRGs in configuration space, as for the CMASS sample.

In our analysis, we do not use the ELGs and QSOs for the following reasons. It is known that for the ELGs sample, the 2-halo model is not able to well describe the power spectrum because the haloes hosting ELGs are undergoing infall motion, and thus imprint filamentary structure onto the clustering signal \citep{2023MNRAS.519.1771O}. 
We do not use the QSOs in DR16 because they are too sparse to measure the cross-correlation between NVSS-SUMSS lens galaxies. Instead, we use the QSO sample from the photometric redshift catalogues provided by Gaia-unWISE samples as explained below.

\subsection{Photometric redshift samples}\label{ssec:photoz_samples}
\subsubsection{2MASS Photometric redshift catalogue}\label{sssec:2mpz}
The 2MASS Photometric Redshift catalogue
(2MPZ: \citejap{2014ApJS..210....9B}) is a galaxy catalogue that covers the local Universe up to $z \approx 0.3$ over most of the sky (except for very low Galactic latitudes). The redshifts of the 2MPZ galaxies were determined by photometric cross-matching with 2MASS, WISE and SuperCOSMOS all-sky samples. 8-band photometry is provided for all the matched galaxies, spanning from the photographic $BRI$, through near-IR $JHK_s$ up to the mid-IR W1 and W2. The error of the photometric redshifts is calibrated through the neural network ANNz \citep{2004PASP..116..345C}, which was trained using the SDSS, 6dFGS, and 2dFGRS counterparts in 2MPZ (over 30\%).
After applying a flux cut of $K_s < 13.9 ({\rm Vega})$ to ensure uniform sky coverage, the final 2MPZ sample includes 
928,352
galaxies with median redshift $z_{\rm med.} = 0.08$ and typical photo-$z$ error $\sigma_{\delta z} = 0.015$, spanning $0 \lesssim z \lesssim 0.3$ (see Figure 12 in \citejap{2014ApJS..210....9B}). We will calculate the angular cross-power spectrum between the NVSS sample and the 2MPZ in harmonic space as the survey mask is simple, following the related study by \cite{Peacock:2018xlz}.

\subsubsection{Quaia: Gaia-unWISE quasars}\label{sssec:quaia}

The Gaia-unWISE quasar catalogue (Quaia) is a highly homogeneous and complete all-sky catalogue of quasars that are derived from ~6.6 million quasar candidates observed by Gaia’s low-resolution blue and red slitless spectrophotometers in the third data release of the space-based {\it Gaia} mission \citep{2024ApJ...964...69S}.

Combined with the unWISE infrared data \citep{2014AJ....147..108L} to improve the sample fidelity, 1,295,502 objects are selected in the original catalogue to the magnitude limit $G < 20.5$, satisfying (1) the number of contaminants such as stars and non-quasar galaxies is reduced by nearly 4$\times$ compared with those based on proper motions and Gaia and unWISE colours, and (2) the number of the objects whose redshifts are ill-determined of the order of $|\Delta z/(1+z)| > 0.2$ are suppressed below $6\%$ with a $k$-nearest neighbour model of colours and redshifts trained through cross-matching with the SDSS DR16 QSOs quasar samples with high-precision spectroscopic redshifts. The median redshift is $z_{\rm med.} = 1.67$.

The all-sky selection function includes astronomical systematics that can reduce the quasar number density: the dust extinction, stellar
density, and the survey pattern of the Gaia satellite (see Sec. 3.3 in \citejap{2024ApJ...964...69S}). Making use of the selection function, the Gaia-unWISE catalogue suits cosmological analyses. For our purpose of measuring the angular cross-power spectra between the NVSS-SUMSS radio samples, we describe the further detail of the selection function in App.~\ref{app:sample_cut_quaia}. We summarise the properties of the reference samples in Table.~\ref{tab:catalogue_property}.

\begin{table}
\begin{center}
\caption{The property of the reference redshift samples and the NVSS-SUMSS radio samples. The highest number of galaxies is in the LOWZ-CMASS samples.} 
\begin{tabular}{ccc}
\hline
Catalogue & \# of samples & range of redshifts \\ \hline \hline
2MPZ & 928,352 & [0.,0.3] \\
LOWZ-CMASS
North & 953,255 & [0.,1.00] \\
LOWZ-CMASS
South & 372,601 & [0.,1.13] \\ 
eBOSS LRG North & 107,500 & [0.60,1.00] \\ 
eBOSS LRG South & 67,316 & [0.60,1.00]\\ 
Gaia-unWISE & 1,295,502 & [0.80,4.]\\ \hline
NVSS+SUMSS &835,552& - \\

\hline \hline
\end{tabular}

\label{tab:catalogue_property}
\end{center}
\end{table}

\subsection{\textit{Planck} CMB lensing}\label{ssec:Planck_CMBlensing}
We employ the CMB lensing map reconstructed with the Planck PR4 data release \citep{Planck:2020olo}. 
The PR4 analysis exploits the improved low-level data processing of NPIPE, accomplishing $20\%$ improvement in signal-to-noise at all scales compared to PR3 analysis.
The details of the study are summarised in \cite{Carron:2022eyg}. 

\section{CMB lensing tomography with the reconstructed NVSS-SUMSS kernel}\label{sec:CMBlensing_tomography_rec_NVSS-SUMSS_kernel}
We estimate the correlation functions in configuration space and angular power spectra in harmonic space depending on the dataset. We employ configuration-space measurement of the cross-correlations between LOWZ-CMASS, eBOSS LRGs, and the NVSS-SUMSS galaxies, where the complicated survey masks are well characterised.
We process the reconstruction of the kernel function of the NVSS-SUMSS galaxies following the narrow redshift slice formulated in Sec.~\ref{sssec:narrow_z_slices}. The noise effects are suppressed  by combining the random catalogues. The random catalogues for the LOWZ-CMASS and eBOSS LRG samples are publicly available, while we create a random catalogue for NVSS-SUMSS sample by uniformly distributing $10^7$ point sources in (RA,DEC) over the sky coverage of NVSS-SUMSS. Note that the information for the line-of-sight distribution of the radio samples cannot be taken into account for this random catalogue since the spectroscopic/photometric redshifts are not measured for the entire radio sample.

In harmonic space, we measure the cross-correlation between the NVSS-SUMSS radio sample with the photometric redshift samples from 2MPZ and Gaia-unWISE following the reconstruction procedure for the broad redshift slice formulated in Sec.~\ref{sssec:broad_z_slices}. We subtract shot noise from the measured power spectra. We derive a method that simultaneously fits the estimated kernel function and the cross-noises given the model of the matter-power spectrum.

Throughout these reconstructions of the kernel function, we primarily select the range of scales of separation or multipole to ensure the scale independence of the reconstructed kernel function, followed by the conditions that reduce other systematics. We supply a way to select the scales in order to mitigate the systematics in App.~\ref{app:cross_noise_in_detail} and ~\ref{app:sample_cut_quaia}.

\subsection{Binning of redshifts for reference samples}\label{ssec:z_bins_of_refs}
\begin{figure*}
 \centering
\includegraphics[width=16.0cm]{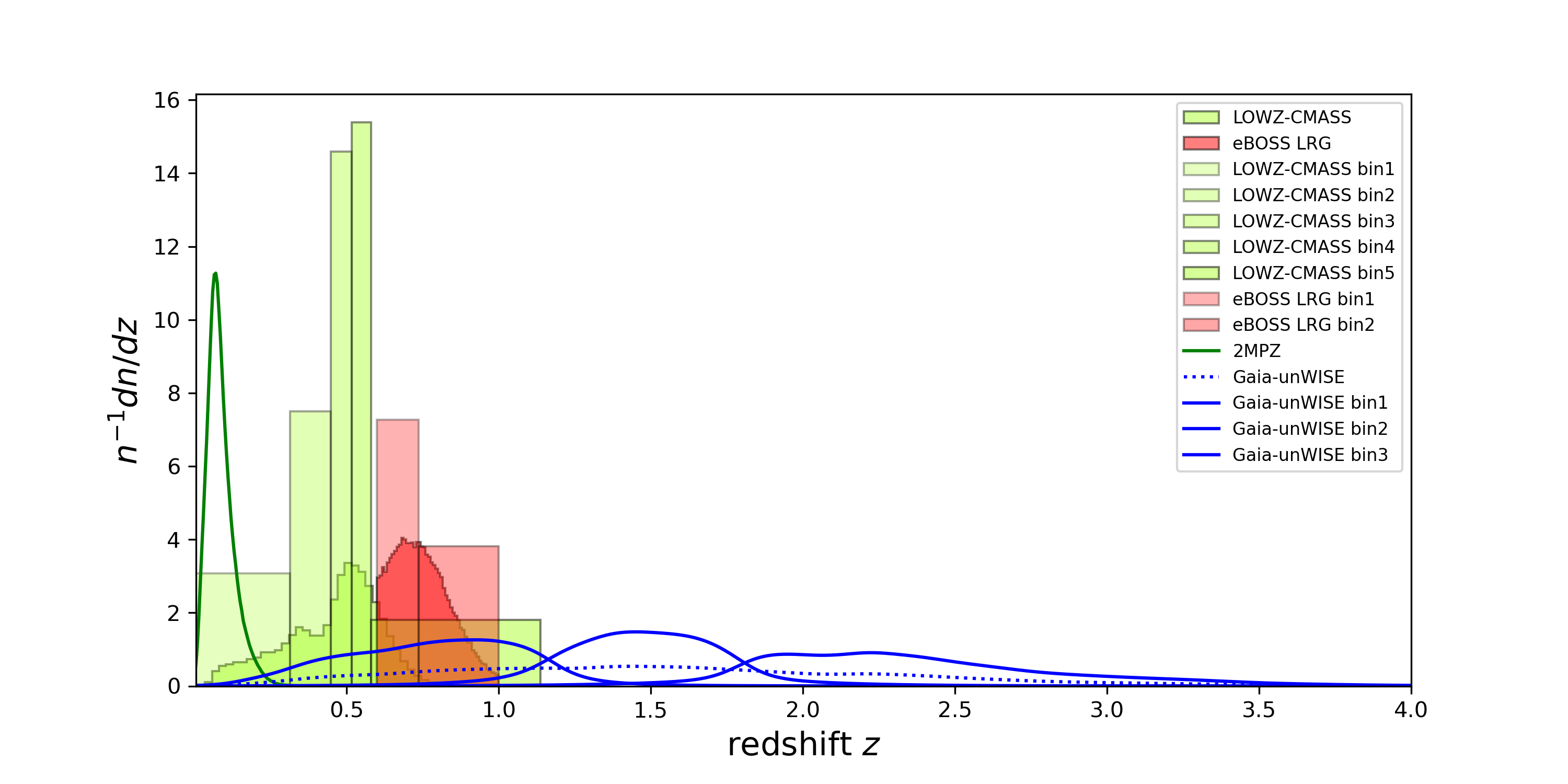}
  \caption{The probability density of the reference samples in redshifts, showing the binning of redshifts of tracers for the reconstruction of the distribution of lensing galaxies. 
  We apply the narrow redshift slices for the LOWZ-CMASS samples and eBOSS DR16 LRGs, whereas the broad redshift slices for 2MPZ and Gaia-unWISE QSOs.}
    \label{fig:redshift_dist_unbinned_binned}
\end{figure*}

We define the bins of redshifts that are utilised for reconstructing the NVSS-SUMSS kernel function. 
We divide the reference redshift samples into the redshift bins to trace the distribution as precisely as possible, whereas the widths should be set so as to retain a high signal-to-noise ratio of measurements while mitigating the RSD effect. Given each reference sample, we employ equipartition of the number of galaxies in each bin. We thus assign a single bin for 2MPZ, 5 bins for LOWZ-CMASS, 2 bins for eBOSS LRGs, and 3 bins for Gaia-unWISE, in total 11 bins spanning $0 \lesssim z \lesssim 3$. The mean redshifts, the widths of the bins, and the projection lengths of the bins are summarised in Table.~\ref{tab:def_of_zslices}. The approximated redshift distributions of the bins are shown in Fig.~\ref{fig:redshift_dist_unbinned_binned}.
We find that the fifth bin in the LOWZ-CMASS and the second bin in eBOSS DR16 LRGs exhibit inhomogeneity within the bins, causing the low amplitudes in the redshift distribution function. Indeed, in Fig.~\ref{fig:redshift_dist_unbinned_binned}, the actual distributions in the last bins for each LOWZ-CMASS and eBOSS LBGs sharply decay within the bin widths. We will not use the last bins of CMASS-LOWZ and eBOSS LRG for the NVSS-SUMSS kernel because these bins are not well approximated by a constant density, which can cause underestimation of reconstructed kernel functions. We will compare the reconstructed NVSS-SUMSS kernel function without the two bins and the ones measured from the two bins in Fig.~\ref{fig:reconstructed_Kg_and_KgxKk_of_NVSS-SUMSS}, and argue that these two bins indeed cause the underestimation of given a model of the kernel function. 

\begin{table}
\begin{center}
\caption{The values of the redshift bin widths and the projection lengths. Photometric samples do not refer to projection lengths, alternatively taking the full ranges of their redshift distributions in projection integrals. The bin widths for photometric samples are set to be twice rms in the sample redshifts.}
\begin{tabular}{cccc}
\hline 
bin name & $z_{\rm mean}$ & $\Delta z$ & $\pi_{\rm max}\;[h^{-1}{\rm Mpc}]$  \\ \hline \hline

2MPZ  & 0.086 & 0.225 & -  \\
LOWZ-CMASS 
bin1 & 0.151 & 0.325 & 904  \\
LOWZ-CMASS 
bin2  & 0.381 & 0.133 & 324  \\
LOWZ-CMASS 
bin3  & 0.482 & 0.0685 & 156  \\
LOWZ-CMASS 
bin4  & 0.548 & 0.0649 &  142  \\
LOWZ-CMASS 
bin5 & 0.858 & 0.555 & 1016\\
eBOSS LRG bin1  &0.668 & 0.137 & 281\\
eBOSS LRG bin2  &0.868 & 0.262 & 475 \\
Gaia-unWISE bin1 & 0.789 & 2$\times$ 0.30 & - \\
Gaia-unWISE bin2 & 1.476 & 2$\times$ 0.29 & - \\
Gaia-unWISE bin3 & 2.348 & 2$\times$ 0.51 & - \\

\hline \hline
\end{tabular}
\label{tab:def_of_zslices}
\end{center}
\end{table}

\subsection{Estimators of two-point correlation function}\label{ssec:est_of_2pt_corr_func}

\subsubsection{Configuration-space correlations}\label{sssec:est_config_sp_corr}
We calculate the configuration-space two-point correlation functions with the Landy-Szalay estimator \citep{1993ApJ...412...64L},
\begin{equation}\label{eq:LS_est}
    w_{AB}(\theta) \equiv \frac{DD_{AB}-DR_{AB}-RD_{AB}+RR_{AB}}{RR_{AB}}\,,
\end{equation}
where $DD$ or $DR$ represent the normalised number of pairs separated within a certain range about an angular radius $\theta$. 
The Landy-Szalay estimator is free from biases due to the finite number of random samples and is known to be the most accurate estimator in the literature. 
We calculate the Landy-Szalay estimator for LOWZ-CMASS, eBOSS DR16 LRGs, and NVSS-SUMSS galaxies, over the sky combined with the north and the south of the SDSS BOSS-eBOSS survey window. The covariance of the measurements is estimated by the bootstrap method processed with 1,000 resamplings from 192 sub-regions divided into approximately equal areas. We calculate the Landy-Szalay estimator using the publicly available package \texttt{TreeCorr}
\citep{2015ascl.soft08007J} with the brute-force computation of the Landy-Szalay estimator i.e. taking $\texttt{binslop}=0$\footnote{The two-point correlation function without the  $\texttt{binslop}=0$ configuration can be inaccurate: cf. \url{https://www2.physik.uni-bielefeld.de/fileadmin/user_upload/theory_e6/Master_Theses/MA_MarianBiermann.pdf}}. 

\subsubsection{Harmonic-space angular power spectra} \label{sssec:est_harmo_sp_corr}
We measure the angular power spectra with a fast numerical transformation of spherical harmonics by using HEALPix \citep{Gorski:2004by}. We divide the sky into $N_{\rm pix}$ pixels where $N_{\rm pix} = 12{N^2_{\rm side}}$ with $N_{\rm side}=1024$, and thus the area in the unit pixel is about $11.8\ {\rm arcmin}^2$.

Let us consider the geometry of surveys. All surveys define only a limited region of the sky over which their properties can be considered statistically homogeneous, so that we need to infer information about the entire sky from the data within limited masks. In configuration-space analysis, masks are straightforwardly treated as a selection of sub-samples out of the full sample; but masks in a harmonic-space analysis need to  be de-convolved in order to extract the information of interest. Here we briefly introduce the masking effect in harmonic space in a simple way as follows. Let us think of the masked field as 
\begin{equation}\label{eq:masked_field_A}
    \delta^W_A = W({\bf n}; \Omega_A)\delta_A\,,
\end{equation}
where the mask is defined as
\begin{align}
W({\bf n}; \Omega) = 
\begin{cases}
1 \ ({\bf n} \in \Omega)\\
0 \ ({\bf n} \notin \Omega)\,.
\end{cases}
\label{eq:window_func}
\end{align}
We shall introduce $f_{\rm sky} \equiv \Omega/4\pi$ to denote the unmasked fraction of the all-sky solid angle. For a given galaxy sample $\delta_g$ is estimated by
\begin{equation}
    \delta_g({\bf n}) \equiv \frac{n_g({\bf n})}{\bar{n}_g \cdot s({\bf n})}-1\,,
\end{equation}
where $\bar{n}_g$ is the average number density of a galaxy sample and $s({\bf n})$ denotes the effective fraction of a tracer galaxy given a specific selection function.

While in configuration space the mask is multiplicative to the observables, in Fourier space this multiplication creates convolution, ending up with a mode mixing between different multipoles $\ell$. This problem is circumvented as far as considering that the angle $2\pi/\ell$ is smaller than the size of the masks. In this case, mode couplings are suppressed obtaining a simple formula so-called 'Pseudo-$C(\ell)$' estimator \citep{Wandelt:2000av,Hivon:2001jp,Gorski:2004by},
\begin{equation}\label{eq:pseudo_cl}
    \hat{C}(\ell) \approx C^{\rm masked}(\ell)\frac{\left < s \right >_{\Omega}}{\left < s^2 \right >_{\Omega}}\,,
\end{equation}
where $\left< x\right>_{\Omega}$ denotes the average of $x({\bf n})$ over the unmasked area. In our analysis, we typically focus on the scales of $\mathcal{O}(\ell)=100$ and the size of the mask is of order $10$ deg., therefore the Pseudo-$C(\ell)$ approximation, Eq.~\eqref{eq:pseudo_cl}, is still valid. A more detailed analysis may be needed by using the original treatment of masks depending on the multipoles $\ell$ in \cite{1973ApJ...185..413P,1973ApJ...185..757H}, and the 'unified pseudo-$C(\ell)$ framework' of \cite{2019MNRAS.484.4127A}. Throughout our measurement, we simply apply the simple pseudo-$C(\ell)$ estimator since the scales of interest are small in comparison with the masks.

We now describe the geometry of the masks for the samples used here. The mask for the 2MPZ sample is selected from the mask for WISE$\times$SuperCOSMOS galaxy catalogue (see \citejap{2016ApJS..225....5B} for details of the construction of the WI×SC mask) as a conservative choice, following the analysis of CMB lensing tomography presented in \cite{Peacock:2018xlz}. 
\begin{figure}
    \centering
    \begin{minipage}[t]{0.48\textwidth}
        \centering
        \includegraphics[width=7.5cm]{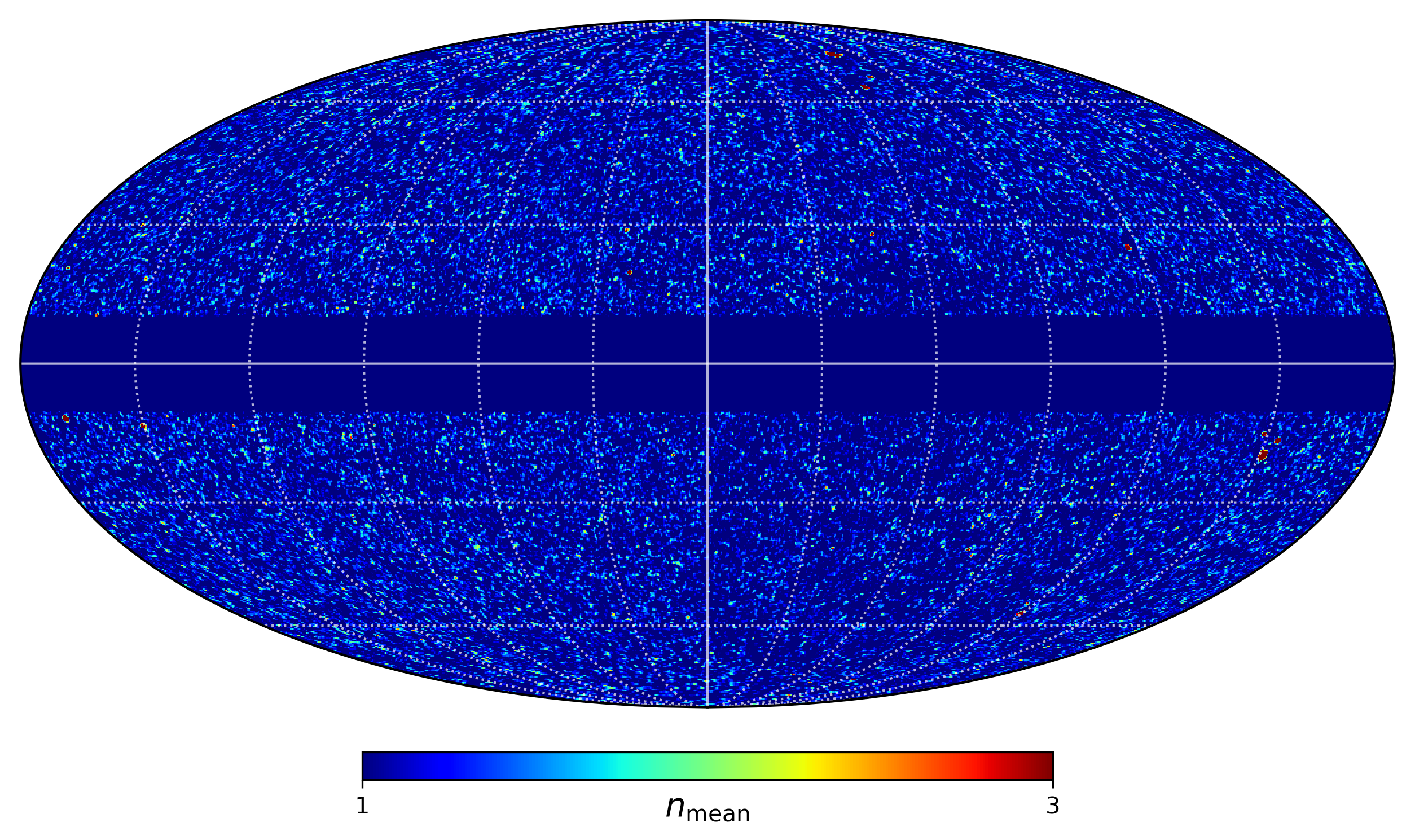}
    \end{minipage}%
    \hfill
    \begin{minipage}[t]{0.48\textwidth}
        \centering
        \includegraphics[width=7.5cm]{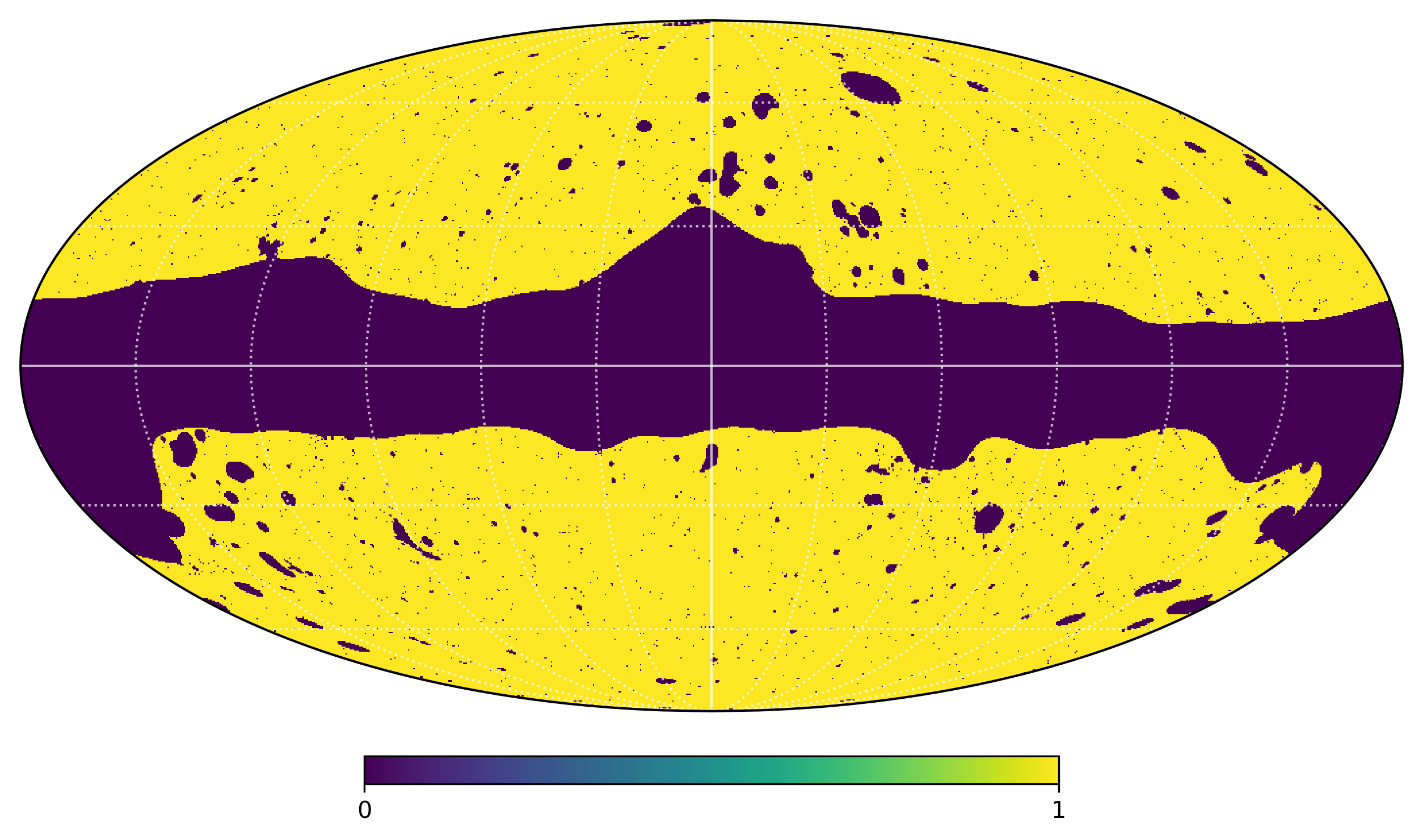}
        \caption{Illustrations of sky maps with a Mollweide projection in galactic coordinates. The top figure shows the NVSS-SUMSS map smoothed at $0.5^\circ {\rm FWHM}$ spanning between 0 and 3  the mean surface number density $n_{\rm mean} = 24.4\,{\rm deg^{-2}}$, with the masking of radio emission from the galactic plane at $|b| \leq 10^\circ$. Note that there are some spots in which fewer galaxies are observed in the Southern celestial hemisphere, whereas the mean surface number density is almost uniform. The bottom figure shows the CMB lensing mask, which already includes the radio $|b| \leq 10^\circ$ masked region.}
        \label{fig:mollview_NVSS_SUMSS_PR4}
    \end{minipage}
\end{figure}

The mask for Gaia-unWISE is designed using the selection function that carefully takes into account the effects of  systematics: dust extinction, stellar density, and the survey pattern of the Gaia satellite (see Sec. 3.3 in \citejap{2024ApJ...964...69S}). The mask is constructed using  $N_{\rm side}=64$  ($\sim 1\,{\rm deg./pixel}$).  In the first harmonic analysis of the Gaia-unWISE sample \citep{2023JCAP...11..043A}, the authors chose a threshold in the selection function $s$ as $s>0.5$ and refitted the weight of the selection functions depending on the redshift slices, i.e., the three bins as described in Table.~\ref{tab:def_of_zslices}. We do not follow this approach, but rather validate the choice of the threshold value of the selection function conservatively, using the errors of the reconstruction to see how sensitively the reconstructed kernel function depends on the threshold. We confirm that the threshold choice has little impact on the reconstructed kernel function within the error bars (see  App.~\ref{app:sample_cut_quaia} for more detail). We choose a representative threshold of $s>0.8$ for our main results.

The mask for the NVSS-SUMSS sample is the simplest, consisting just of a cut in  galactic latitude below $|b| = 10^\circ $. The masked map of the NVSS-SUMSS map is shown in the top figure in Fig.~\ref{fig:mollview_NVSS_SUMSS_PR4}. When computing the angular cross-power spectrum between the NVSS-SUMSS sample and \textit{Planck} PR4 convergence map, the mask for the NVSS-SUMSS sample is completely contained within the mask of the {\it Planck} PR4 lensing map.

\subsection{Estimators of kernel function}\label{ssec:est_of_kernel_func}
We now derive the estimator and the error of the reconstructed kernel function. Assuming independence of the angular scales in the reconstructed kernel, there are individually measured kernel functions at the location of the bins in angle or multipole i.e. $M_\theta$ or, $\ell_{\rm bin}$, respectively. In what follows, we derive the estimator separately for configuration-space or harmonic-space measurements. 

\subsubsection{Estimator in configuration space}
\label{sssec:est_config-sp_analysis}
The kernel function can be, in general, denoted as $K^{\rm rec.}_{g}(z_i,\theta_I)$.
We combine the measured lensing kernel for each angular bin into the averaged estimator,
\begin{equation}\label{eq:kernel_lens_avg}
    K_{g}(z_i) \equiv \frac{1}{M_\theta}\sum^{M_\theta}_{I=1}K^{\rm rec.}_{g}(z_i,\theta_I)\,.
\end{equation}
where $I$ denotes the index of the bin of the angle. The error budget of $K_g(z_i)$ is obtained as
\begin{align}\label{eq:sigma2_K}
    \sigma^2_K(z_i) = {}^T Y \cdot  {\rm Tr}(\mathbb{COV}) \cdot Y\,,
\end{align}
where $Y \equiv {}^T(\partial K_g/\partial w, \partial K_g/\partial w')$ and the covariance matrix ${\mathbb{COV}}$ is defined as 
\begin{equation}\label{eq:covariance_ww'IJ}
    \mathbb{COV}_{w w',IJ}(z_i) \equiv \frac{M-1}{M}\sum^M_{k=1}{X^{(k)}_{w,I}X^{(k)}_{w',J}}\,. 
\end{equation}
Here $X^{(k)}_{w,I} \equiv w^{(k)}(z_i,\theta_I)-\hat{w}(z_i,\theta_I)$ where $w^{(k)}(z_i,\theta_I)$ the angular correlation measurement at separation $\theta_I$ on the $k$-th bootstrap subsample, where we have $M=192$ subsamples. $\hat{w}$ is the averaged correlation function over all the bootstrap sub-samples,
\begin{equation}\label{eq:w_avg}
    \hat{w}(z_i,\theta_I) = \frac{1}{M}\sum^M_{k=1}w^{(k)}(z_i,\theta_I)\,.
\end{equation}
Note that the trace in Eq.~\eqref{eq:sigma2_K} is taken for the indices of the bins of the angle $I$ and $J$.

\subsubsection{Estimator in harmonic space}\label{sssec:est_harmo-sp_analysis}
We derive an estimator of the kernel function dealing with shot noises in harmonic analysis, particularly focusing on the angular cross-power spectra of two galaxy catalogues. In an ideal case where the two samples are generated by completely different random processes, there is no shot noise correlation between them. But in practical cases there are some sample overlaps, and two samples are then not statistically independent. Under this situation, the pseudo-$C(\ell)$ estimator denoted $\hat{C}(\ell)$ between the sample $A$ and the sample $B$ is given as
\begin{equation}\label{eq:pseudo_cl_w_crossnoise}
    \hat{C}_{AB}(\ell) = C_{AB}(\ell) + \frac{n_{AB}}{n_A n_B}\,,
\end{equation}
where $n_A$, $n_B$, and $n_{AB}$ denote the number of the sample of $A$, $B$, and the objects mutual both in $A$ and $B$ per unit solid angle, respectively. Note that $n_{AB} = n_A$ if $A=B$, otherwise $n_{AB} < {\rm min}(n_A,n_B)$. 
$n_A$ is determined through $\smash{n_A = n^{\rm masked}_A}$, where $\smash{n^{\rm masked}_A}$ denotes the number density of the sample $A$ in the masked region. 

We subtract the shot noise $1/n_A$ from the auto-power spectra of the sample $A$ without further processing. To subtract the shot noise in the angular cross-power spectrum between the reference and the unknown samples, namely the cross-noise,  we fit the amplitude of the noise based on  Eq.~\eqref{eq:pseudo_cl_w_crossnoise}. Ideally, the cross-noise is determined if the number of mutual objects in both samples is identified. In practice, we simply fit the angular cross-power spectrum with the following model and derive the amplitude of cross-noise as
\begin{align}\label{eq:cross_noise_model}
    \frac{\hat{C}_{rr}(\ell,z_i)}{E[C(\ell,z_i)]} &= b^2_r(z_i) + \frac{1}{n_r(z_i) E[C(\ell,z_i)]}\,,\\
    \frac{\hat{C}_{gr}(\ell,z_i)}{E[C(\ell,z_i)]} &= b_r(z_i)K_g(z_i) + \frac{a_{gr}(z_i)}{n_r(z_i) E[C(\ell,z_i)]}\,,
\end{align}
by taking $a_{gr}$ to be an additional free parameter.
One can confirm whether $a_{gr}$ obtained from a fitting is consistent with the surface density of the mutual objects between the $g$ and $r$ samples by using  $n_{gr} = a_{gr} n_g$, 
where $n_{gr}$ denotes the number density per unit solid angle of the mutual objects that are in both the NVSS-SUMSS sample and the reference sample i.e. 2MPZ and Gaia-unWISE QSOs. This is because one can derive $n_{gr}$ by cross-matching the samples and identifying the mutual objects. 
The assumption we take here is that the kernel function should be scale-independent in multipole moments. We fit the measured angular cross-power spectra in each redshift bin and simultaneously obtain $b^2_r$, $b_rK_g$ and $a_{gr}$ at $z = z_i$. 

In Table~\ref{tab:cross_noise}, we summarise the best-fitting values of $b^2_r$, $b_rK_g$ and $a_{gr}$ for the reconstruction from 2MPZ and Gaia-unWISE QSOs, respectively. Note that the best-fit values of $a_{gr}$ are well-determined and less than unity, showing the further stronger posterior constraints on the prior range $0 < a_{gr}<1$. One may consider the cross-noise estimate via cross-matching either between 2MPZ and NVSS-SUMSS or Gaia-unWISE quasar NVSS-SUMSS within a certain sky region clean from astrophysical systematics. We leave this further investigation for future work.

To analyse the scale where the linear bias model in Eq.~\eqref{eq:delta_G_3D} holds, we cut the range of the angular scales and the multipole moments as follows. 
We consider only $\theta \in [0.3^\circ, 1^\circ]$ when computing the cross-correlation between the NVSS-SUMSS sample and the spectroscopic samples: LOWZ-CMASS and eBOSS LRGs. The lower cut is determined by the previous studies on the measurements of galaxy clustering for LOWZ-CMASS and LRGs which ensures that the linear bias model is applicable. The lower cut may also be useful in mitigating unknown systematics in cross-correlations due to multiple images corresponding to a single double radio galaxy in the radio catalogues. The upper cut is designed to mitigate the effect of a significant amount of the unknown non-Gaussian component (see \citejap{2010ApJ...717L..17X}, \citejap{2010A&A...520A.101H}, \citejap{2016A&A...591A.135C}). 
We set $20 \leq \ell \leq 250$ for 2MPZ following the previous analysis in \cite{Peacock:2018xlz} that also applied the HALOFIT nonlinear model and a scale-independent bias. For Gaia-unWISE we adopt $180 \leq \ell \leq 540$. We cut the largest multipole to $\ell = 540$ so that the linear approximation of the density contrast is applicable. On the other hand, we cut the smallest multipole when analysing the un-WISE QSOs, so as to avoid the systematic leakage of dust emission (see  \citejap{2023JCAP...11..043A}). 
In App.~\ref{app:cross_noise_in_detail}, we expand the analysis to other choices of $\ell_{\rm min.}$, showing how the results depend on the choice of the lowest multipoles, concluding that for $\ell \geq 100$ the amplitude of the reconstructed kernel function is not biased by  dust contamination. We calculate $K_g$ from the fitted values of $b^2_r$ and $b_r K_g$. The error $\sigma^2_K(z_i)$ is calculated from the standard deviation of the MCMC samples with the posterior probability derived from the MCMC chain provided by \texttt{emcee}\footnote{\url{https://emcee.readthedocs.io/en/stable/}}. The best-fit values of $b^2_r$, $b_r K_g$, $a_{gr}$ are shown in Table.~\ref{tab:cross_noise}.

\def\spacer{\phantom{_A}}

\begin{table*}
\begin{center}
\caption{The fitted values and the errors of $b_rK_g$ and $a_{gr}$ for 2MPZ $(20\leq \ell \leq 250)$ and Gaia-unWISE $(180\leq \ell \leq 540)$. The priors are uniform: $b_r \in [0.,10.]$, $b_rK_g \in [0.,10.]$ and $a_{gr} \in [0.,1.]$.
The values indicate the median, 0.16 and 0.84 percentiles for the lower and the upper bound, respectively. The posterior distributions are displayed in App.~\ref{app:cross_noise_in_detail}.}
\begin{tabular}{ccccc}
\hline
sample & $z_i$ &$b_r$ &$b_rK_g$& $a_{gr}$ \\ \hline \hline
2MPZ & 0.074 &$0.99^{+0.007}_{-0.007}$
&$0.60^{+0.05}_{-0.06}$ & $0.46^{+0.28}_{-0.26\spacer}$\\
Gaia-unWISE bin1 & 0.789 &$1.42^{+0.22}_{-0.22}$ & $2.29^{+0.75}_{-0.74}$ & $0.37^{+0.12}_{-0.12\spacer}$ \\
Gaia-unWISE bin2 & 1.476 & $2.30^{+0.56}_{-0.56}$
&$2.69^{+1.02}_{-1.08}$ & $0.19^{+0.11}_{-0.10\spacer}$ \\
Gaia-unWISE bin3 & 2.348 &$3.34^{+1.73}_{-1.73}$& $1.27^{+1.25}_{-0.87}$ & $0.38^{+0.08}_{-0.09\spacer}$ \\
\hline \hline
\end{tabular}
\label{tab:cross_noise}
\end{center}
\end{table*}

\subsection{Results of CMB lensing tomography}\label{ssec:results_of_CMB lensing_tomography}
We are now in a position to carry out the tomographic analysis of the CMB lensing signal, using the reconstructed NVSS-SUMSS kernel function. The main output of this analysis will be an estimate of the amplitude of the matter density fluctuations. We measure the angular cross-power spectrum between NVSS-SUMSS galaxies and the \textit{Planck} PR4 CMB lensing convergence by using the pseudo-$C(\ell)$ method as described in Sec.~\ref{sssec:est_harmo_sp_corr}. Then we estimate $\sigma_8$ via a maximum likelihood analysis, fitting the measured $C_{g\kappa}(\ell)$ by the analytical method described in Sec.~\ref{ssec:CMB lensing_tomography_with_rec_Ku}. To clarify the uncertainty from the highest-redshift residual part in estimating $\sigma_8$ that we introduced in Eq. \eqref{eq:alpha_sigma8_estimator}, we will consider two models for the NVSS-SUMSS kernel function that take into account nuisance parameters in estimating $\sigma_8$. The first model in Sec.~\ref{sssec:model1_with_constant_Kres} provides a minimum model with two constant parameters, whereas the second model in Sec.~\ref{sssec:model2_with_2lognormal_K} consists of four nuisance parameters in a continuous fitting function of redshift. We will argue that using the latter model for the NVSS-SUMSS kernel function is preferred in $\chi^2$ statistics, successfully treating uncertainty from the residual signal in estimating $\sigma_8$. After marginalising over the uncertainty of the nuisance parameters, we will show that the estimation of $\sigma_8$ can be reliable given our choice of model and dataset, quantitatively assessing the relation between the deepest redshift that is decomposable as a tomographic bin and the noise in the measurement of $C_{g\kappa}$ (see Fig.~\ref{fig:separable_redshifts_given_noise_from_cgk_PR4_NVSS-SUMSS}). After these steps, we obtain the constraint $\sigma_8 = 0.86^{+0.12}_{-0.09}$.

\subsubsection{Model with a constant $K^{\rm res.}_g$}\label{sssec:model1_with_constant_Kres}
As we show in Sec.~\ref{ssec:CMB lensing_tomography_with_rec_Ku}, $C_{g\kappa}(\ell)$ is computed by combining two parts: the summation of the reconstructed part $\Sigma_{g\kappa}$ and the high-redshift residual part $S_{g\kappa}$. For the ideal case where the reference sample completely covers the redshift distribution of the lens galaxies, $S_{g\kappa}$ has no systematic effect on the measured $C_{g\kappa}(\ell)$ as $\Sigma_{g\kappa} \gg S_{g\kappa}$. In our case, however, this ideal property is not guaranteed: the NVSS-SUMSS can extend to $z > 2.3$ where few reference samples exist, and thus we need to estimate $S_{g\kappa}$ as well as $\sigma_8$ taking this into account.

We introduce a constant $K^{\rm res.}_g$ model for the angular cross-power spectrum $C_{g\kappa}(\ell)$ as
\begin{equation}\label{eq:cgk_model_with_constant_Kres}
    C^{\rm model}_{g\kappa}(\ell) = \frac{\sigma_{8,{\rm est.}}}{\sigma_{8,{\rm fid.}}}\Sigma\sum_i\omega_i(\ell) + \left(\frac{\sigma_{8,{\rm est.}}}{\sigma_{8,{\rm fid.}}} \right)^2 S_\kappa(\ell)\,,
\end{equation}
where
\begin{align}
    \omega_i(\ell) &\equiv \frac{K_\kappa(z_i)\delta z_i}{\chi^2(z_i)c/H(z_i)}P_m\left(\frac{\ell+1/2}{\chi(z_i)},\,z_i\right)\,,\nonumber \\
    S_\kappa(\ell) &\equiv K \int^{z_*}_{{z_{\rm max}}}{ {\rm d}z\frac{K_\kappa(z)}{\chi^2(z)c/H(z)} P_m\left(\frac{\ell+1/2}{\chi(z)},z\right)}\,,\nonumber \\
    K &= K_g^{\rm res.} = {\rm const.}
\end{align}
Note that $i$ spans all the redshift bins where the NVSS-SUMSS kernel function is reconstructed (see the detail about the binning in Fig.~\ref{fig:redshift_dist_unbinned_binned} and Table.~\ref{tab:def_of_zslices}). The first term on the right-hand side in Eq.~\eqref{eq:cgk_model_with_constant_Kres} fits the summation derived from the reconstructed kernel function in each bin by $\Sigma$. The second term in Eq.~\eqref{eq:cgk_model_with_constant_Kres} fits the residual component coming from $z > z_{\rm max.}$ where the reconstruction of the kernel is not performed. Note that the first term in the right-hand side scales as $\sigma_{8,{\rm est.}}/\sigma_{8,{\rm fid.}}$ because $\Sigma$ fits the reconstructed kernel function that is already proportional to $\sigma_{8,{\rm est.}}/\sigma_{8,{\rm fid.}}$ (see the discussion of this property in Sec.~\ref{ssec:clustering_z}).

We adopt a maximum likelihood analysis to estimate the parameters. The log-likelihood of this model is defined as
\begin{align}
    {\rm ln}(\mathcal{L}) \equiv& 
    -\sum_\ell{\left(\frac{(\hat{C}_{g\kappa}(\ell)-C^{\rm model}_{g\kappa}(\ell))^2}{2\hat{\sigma}^2_{g\kappa}(\ell)} \right)} \cr 
    &- \left(\frac{(\Sigma(\ell)
    -\Sigma \sum_i(\omega_i(\ell))}{2\sigma^2_\Sigma(\ell)}\right)\,.
\end{align}
$\sigma_\Sigma(\ell)$ is defined as
\begin{equation}
\sigma^2_\Sigma(\ell) \equiv \sum_i \omega^2_i(\ell)\sigma^2_K(z_i)\,,
\end{equation}
provided that the errors in the kernel function for each bin are statistically independent. Here $\hat{C}_{g\kappa}(\ell)$ and $\hat{\sigma}^2_{g\kappa}(\ell)$ are the measured pseudo angular cross-power spectrum and its standard deviation, respectively.

We show the derived constraints on the parameters with the posterior distributions in Fig.~\ref{fig:contour_constant_Kres}. The best-fit parameters are shown in Table ~\ref{tab:constant_Kres_bestfit}. We find that $\sigma_8$ is degenerate with $K$, which can be understood as follows. $\Sigma$ is determined by the reconstructed kernel function by the second term of the likelihood as well as the amplitude of $C_{g\kappa}$. Then $K$ fits $C_{g\kappa}$ given the value of $\Sigma$. Since $\sigma_8$ and $K$ both both amplify $C_{g\kappa}$, $\sigma_8$ and $K$ have an anti-correlation in their values once $C_{g\kappa}$ is fixed by the measurement. In addition, we also find that the best-fit $\sigma_8$ is smaller. The smaller $\sigma_8$, however, seems not to be cosmological but to be systematics of the NVSS-SUMSS kernel assumed here.
The constant amplitude of $K^{\rm res.}_g$ tends to overestimate the ingredients of $C_{g\kappa}$ from the higher redshift part of the correlation, making the estimated $\sigma_8$ smaller than that of \textit{Planck} 2018 \citep{Planck:2018lbu} given the measured amplitude of $C_{g\kappa}$. This is because the kernel function must decrease as the redshift becomes higher, reflecting the fact that there are fewer samples of the NVSS-SUMSS galaxies at higher redshifts given the flux limit for the entire sample. These results motivate the construction of a better model of the NVSS-SUMSS kernel function to measure $\sigma_8$, characterising the features of the kernel function with which the reconstructed kernel function and the residual kernel function are properly correlated. In the next subsection, we will introduce a candidate for such a model.

\subsubsection{Double lognormal model for the NVSS-NVSS kernel}\label{sssec:model2_with_2lognormal_K}
We introduce a fitting formula of the radio kernel function as
\begin{align}\label{eq:double_lognormal_kernel_model}
    K^{\rm model}_g(z) = &\frac{\sigma_{8,{\rm est.}}}{\sigma_{8,{\rm fid.}}}\Biggl\{(\alpha_1+\alpha_2) \exp\left(-\frac{{\rm ln}^2(z)}{\beta^2_1}\right) \cr
    &+ \alpha_2 \exp\left(-\frac{{\rm ln}^2(z)}{(\beta_1+\beta_2)^2}\right)\Biggr\}\,.
\end{align}
Note that $K^{\rm model}_g(z)$ is proportional to ${\sigma_{8,{\rm est.}}}/{\sigma_{8,{\rm fid.}}}$ as we fit the reconstructed kernel function (see the further discussion on this in Sec.~\ref{sec:CMB-lensing_tomography_with_rec_kernel}). This model tries to capture the following features of the NVSS-SUMSS kernel function, introducing the four parameters $\alpha_{1,2}$ and $\beta_{1,2}$. Firstly, the redshift distribution of the sample converges to zero at $z = 0$. Secondly, the redshift distribution of the NVSS-SUMSS sample is likely to have two distinct peaks, representing the two different populations of radio sources: star-forming galaxies at low redshifts and the AGN at higher redshifts (see  \citejap{2007MNRAS.375..931M} and for other radio samples cf. \citejap{2013MNRAS.436.1084M}).
Finally, the lognormal distribution has a mild decrease of its tail towards higher redshifts, featuring the tail distributions of bright AGN at higher redshifts.  We set the parameter space with the hierarchy of the parameters as $\alpha_{1,2}>0.$ and $\beta_{1,2} >0$. The first term in Eq.~\eqref{eq:double_lognormal_kernel_model} features a larger amplitude with a smaller deviation, tracing the AGN population. The second term in Eq.~\eqref{eq:double_lognormal_kernel_model}, on the other hand, features a smaller amplitude with a wider deviation, following the distribution of the star-forming galaxies.  Then $C_{{\rm radio}\ \kappa}$ is derived as
\begin{equation}
    C_{g\kappa}(\ell) \equiv \frac{\sigma_{8,{\rm est.}}}{\sigma_{8,{\rm fid.}}}\int^{z_*}_{0}{\rm d}z\frac{K^{\rm model}_g(z)K_\kappa(z)}{\chi^2(z)c/H(z)}P_m\left(\frac{\ell+1/2}{\chi(z)},\,z\right)\,.
\end{equation}
We adopt the maximum likelihood estimation of the best-fit parameters with the following log-likelihood,
\begin{align}\label{eq:log_likelihood_kernel}
    {\rm ln}(\mathcal{L}) \equiv 
    &-\sum_\ell\left\{\frac{(\hat{C}_{g\kappa}(\ell)-C^{\rm model}_{g\kappa}(\ell))^2}{2\hat{\sigma}^2_{g\kappa}(\ell)} \right\} \cr
    \qquad &-\sum_i\left\{\frac{(K_g(z_i)-K^{\rm model}_g(z_i))^2}{2\sigma^2_K(z_i)}\right\}\,,
\end{align}
and we proceed with MCMC sampling to obtain the posterior distribution of the parameters. Note that $i$ spans all the redshift bins where the NVSS-SUMSS kernel function is reconstructed (see the detail regarding the binning in Fig.~\ref{fig:redshift_dist_unbinned_binned} and Table.~\ref{tab:def_of_zslices}).

We show the derived constraints on the parameters with the posterior distributions in Fig.~\ref{fig:contour_2lognormal_Kradio}. The best-fit parameters are shown in Table ~\ref{tab:double_lognormal_bestfit}. We find that $\sigma_8$ is consistent with that of \textit{Planck} 2018 obtained from TT-TE-EE correlations in the 2-$\sigma$ credible interval, considering the degeneracy between $\sigma_8$ and $\alpha_1$. The degeneracies between and $\sigma_8$ and $\alpha_2$ and $\beta_{1,2}$ are not significant. Fig.~\ref{fig:reconstructed_Kg_and_KgxKk_of_NVSS-SUMSS} shows the comparison between the reconstructed kernel function and the best-fit model. We find that $\alpha_2$ and $\beta_2$ trace the first two bins from $z=0$.

\begin{figure}
 \centering
  \includegraphics[width=7.5cm]{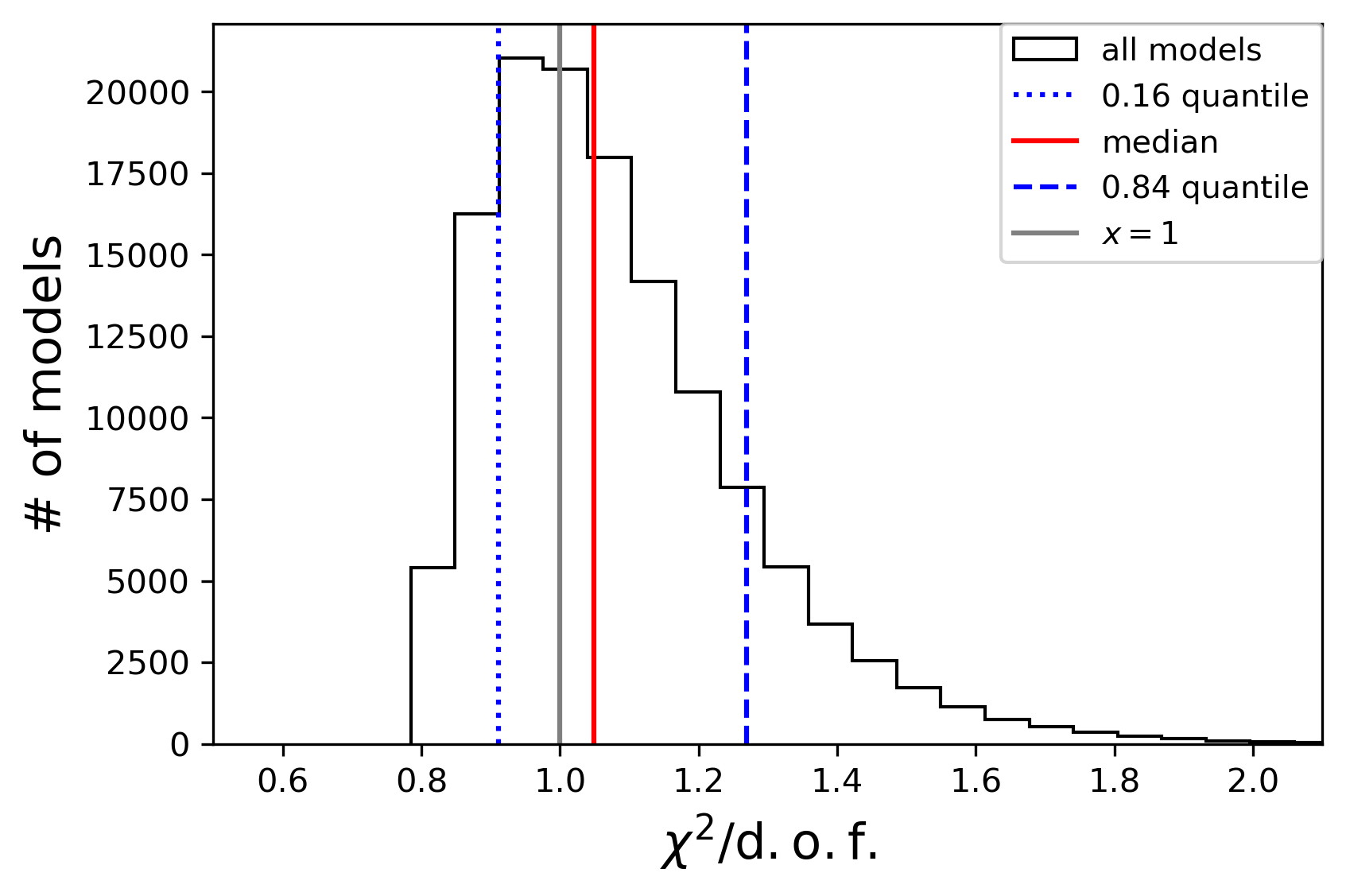}
  \caption{The histogram of $\chi^2/{\rm {d.o.f.}}$. Here we collect 132,600 samples of $(\sigma_8, \alpha_1,\alpha_2,\beta_1, \beta_2)$. The median (red), and the quantile range: 0.16 and 0.84 are shown by the blue dotted and the blue dashed lines, respectively.}
    \label{fig:chi2_per_dof_fitting_kernel}
\end{figure}

Remarkably, the double lognormal model exhibits $\chi^2/{\rm d.o.f.} \approx 0.967$ given $\chi^2 \equiv -2{\rm ln}(\mathcal{L}) $, showing a good match to the data. Fig.~\ref{fig:chi2_per_dof_fitting_kernel} shows the histogram of $\chi^2/{\rm d.o.f.}$ made from all the realisations in MCMC analysis. We find that most of the models are distributed with a small dispersion around unity, indicating that the double lognormal model is a good candidate to explain the data. On the other hand, we find $\chi^2/{\rm d.o.f.} \approx 0.669$ for the constant $K^{\rm res.}_g$ model, indicating that this model seems insufficient to explain the data compared to the double lognormal model. Fig.~\ref{fig:comparison_dgk_rec_vs_measurement} compares the measured angular cross-power spectrum and the theoretical prediction made by the best-fit parameters of the radio kernel functions of the two models.

\begin{figure}
 \centering
  \includegraphics[width=7.5cm]{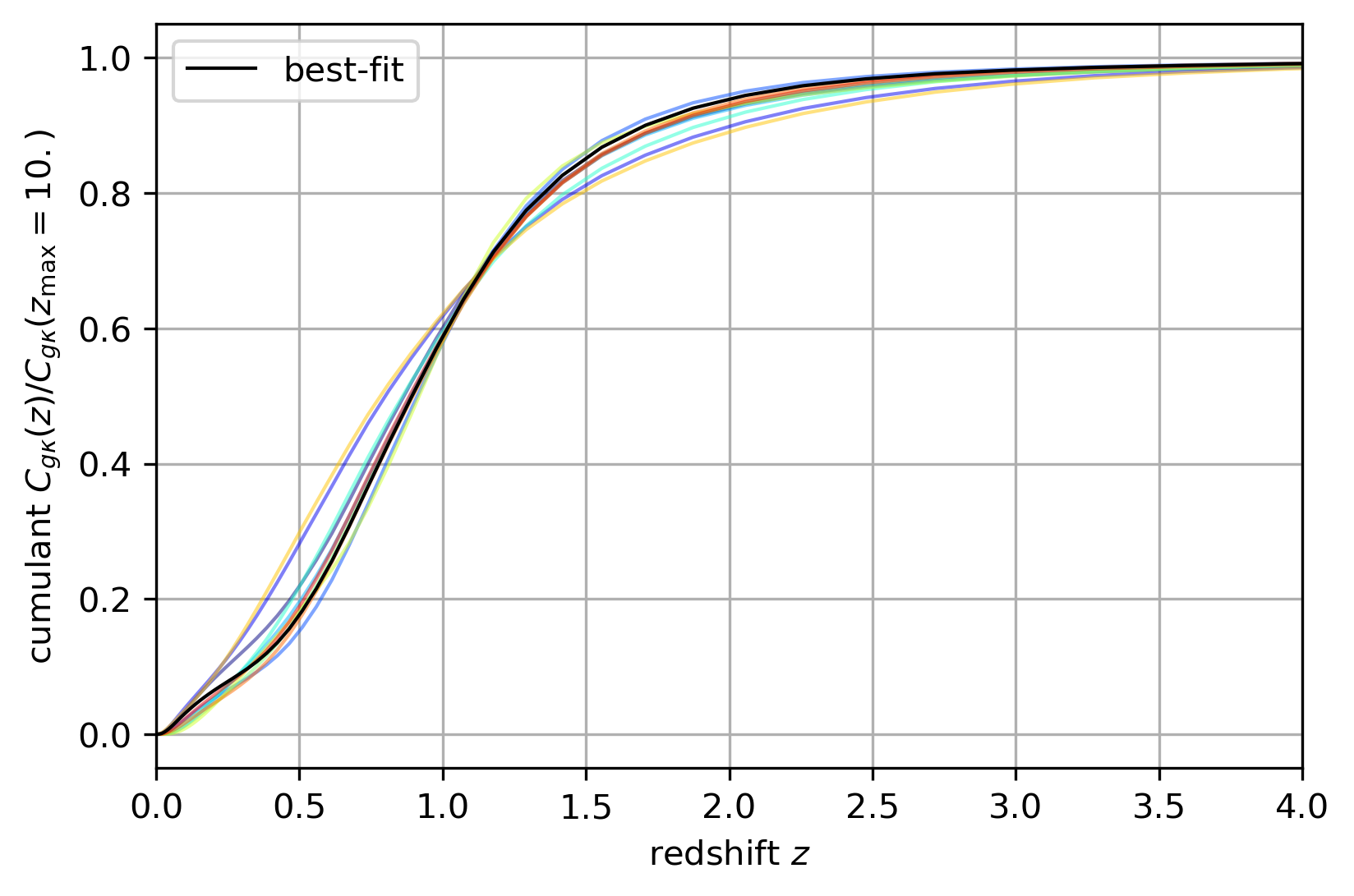}
  \caption{Cumulant of reconstructed $C_{g\kappa}(\ell=225)$. The black line is the model computed with the median. The other coloured lines are 10 random samples drawn from the entire 132,600 MCMC realisations.}
    \label{fig:cumulant_cgk}
\end{figure}
\begin{figure}
 \centering
  \includegraphics[width=7.0cm]{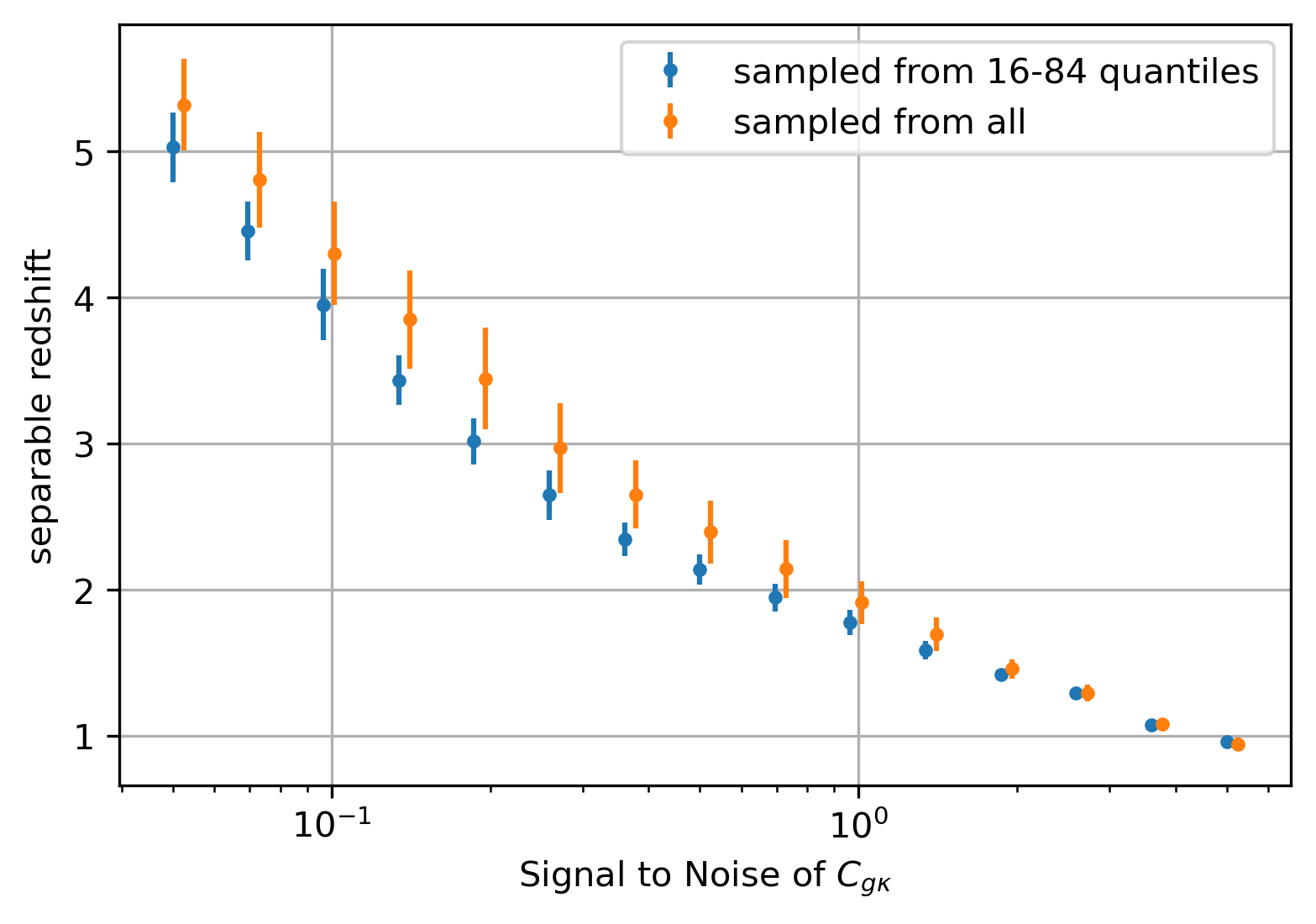}
  \caption{Separable redshifts as a function of signal to noise ratio. The blue and orange points represent the different 10 subsamples of the entire 132,600 MCMC realisations: the blue points were sampled between 0.16-0.84 quartiles of $\chi^2$ distribution, whereas the orange ones were sampled from the entire distribution (see the $\chi^2$ distribution in Fig.~\ref{fig:chi2_per_dof_fitting_kernel})}
    \label{fig:separable_redshifts_given_noise_from_cgk_PR4_NVSS-SUMSS}
\end{figure}

\begin{figure*}
    \vspace{2.0cm}
    \centering
    \begin{minipage}{1.0\textwidth}
        \centering
        \captionof{table}{The best-fit parameters of the constant $K^{\rm res.}_g$ model of the angular cross-power spectrum between the NVSS-SUMSS sample and \textit{Planck} PR4 CMB lensing convergence map. The values of the parameters indicate the median, 0.16 and 0.84 percentiles for the lower and the upper bound, respectively.}
        \begin{tabular}{ccccc}
        \hline
        params. & 
        $\sigma_{8,{\rm est.}}/\sigma_{8,{\rm fid.}}$ & $\Sigma$ & $K$ & $\chi^2/{\rm d.o.f.}$ \\ \hline \hline
        best fit & $0.82^{+0.33}_{-0.18}$ & $1.14^{+0.07}_{-0.07}$ & $0.68^{+0.82}_{-0.53}$ & 12.05/18\\
        \hline \hline
        \end{tabular}
        \label{tab:constant_Kres_bestfit}
    \end{minipage}%
    \hfill
    \begin{minipage}{1.0\textwidth}
        \centering
        \includegraphics[width=\textwidth]{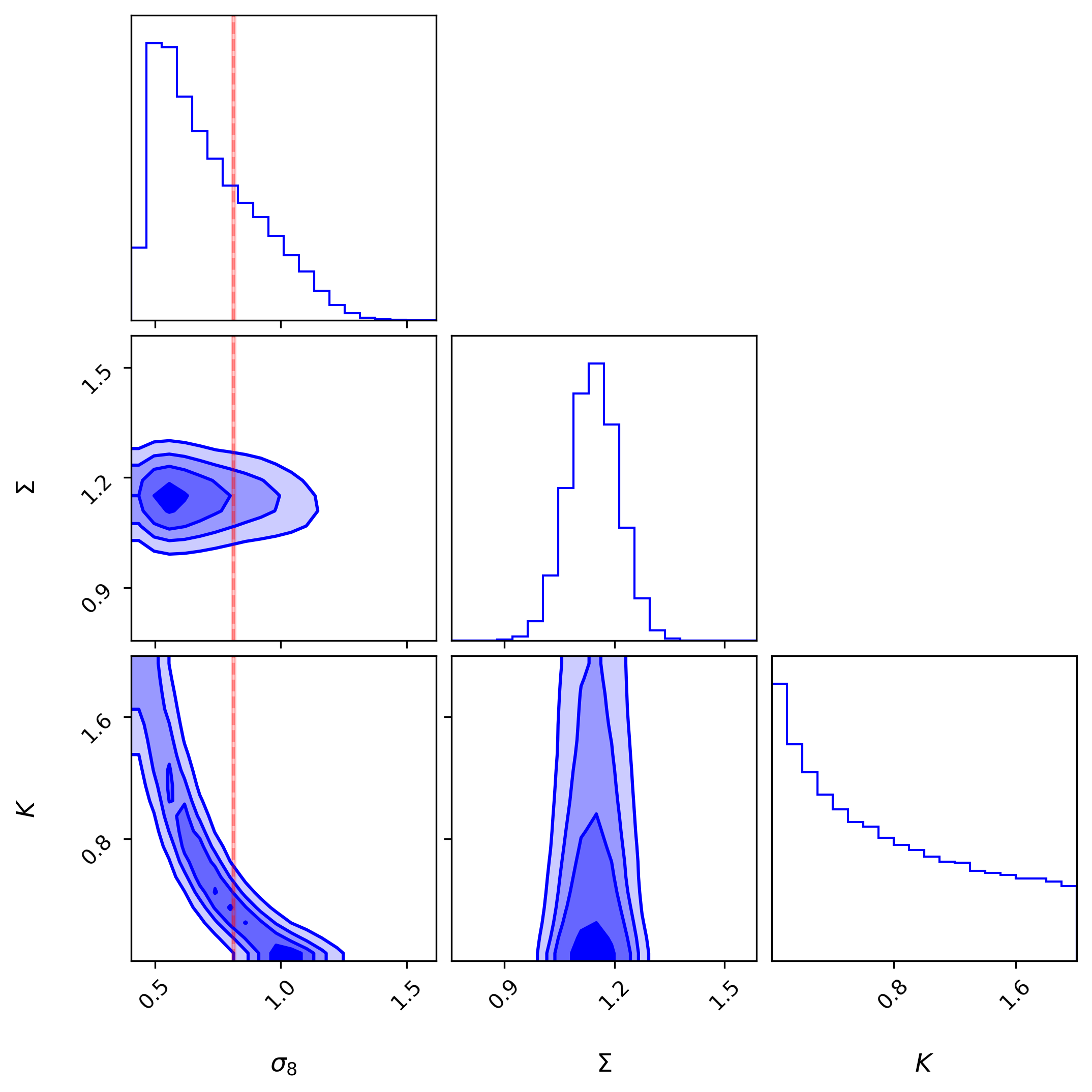}
        \caption{The posterior distributions and the contour maps for the parameters of the constant $K^{\rm res.}_g$ model. The blue contours represent confidence regions, corresponding to 1-$\sigma$ (68\%), 2-$\sigma$ (95\%), or 3-$\sigma$ (99.7\%), and 4-$\sigma$ (99.9973\%) credible intervals, respectively. The red-shaded region shows the 1-$\sigma$ credible interval obtained from \textit{Planck} 2018 TT-TE-EE model \citep{Planck:2018vyg}. The MCMC analysis was configured with 20,000 steps per walker. The first 100 steps were discarded as burn-in, and a thinning factor of 15 was applied, resulting in a total of 132,600 samples.}
        \label{fig:contour_constant_Kres}
    \end{minipage}
\end{figure*}

\begin{figure*}
    \vspace{2.0cm}
    \centering
    \begin{minipage}{1.0\textwidth}
        \centering
        \captionof{table}{The best-fit parameters for the double lognormal model of the NVSS-SUMSS kernel function and $\sigma_8$. The angular cross-power spectrum between the NVSS-SUMSS sample and \textit{Planck} PR4 CMB lensing convergence map. The values of the parameters indicate the median, 0.16 and 0.84 percentiles for the lower and the upper bound, respectively.}
        \begin{tabular}{ccccccc}
        \hline
        params. & 
        $\sigma_{8,{\rm est.}}/\sigma_{8,{\rm fid.}}$ & $\alpha_1$ & $\alpha_2$& $\beta_1$ & $\beta_2$ & $\chi^2/{\rm d.o.f.}$ \\ \hline \hline
        best fit & $1.07^{+0.16}_{-0.12}$ & $1.22^{+0.41}_{-0.37}$ & $0.14^{+0.04}_{-0.03}$  & $0.45^{+0.04}_{-0.03}$ & $1.27^{+0.23}_{-0.17}$ & 15.48/16 \\
        \hline \hline
        \end{tabular}    \label{tab:double_lognormal_bestfit}
    \end{minipage}%
    \hfill
    \begin{minipage}{1.0\textwidth}
        \centering
        \includegraphics[width=\textwidth]{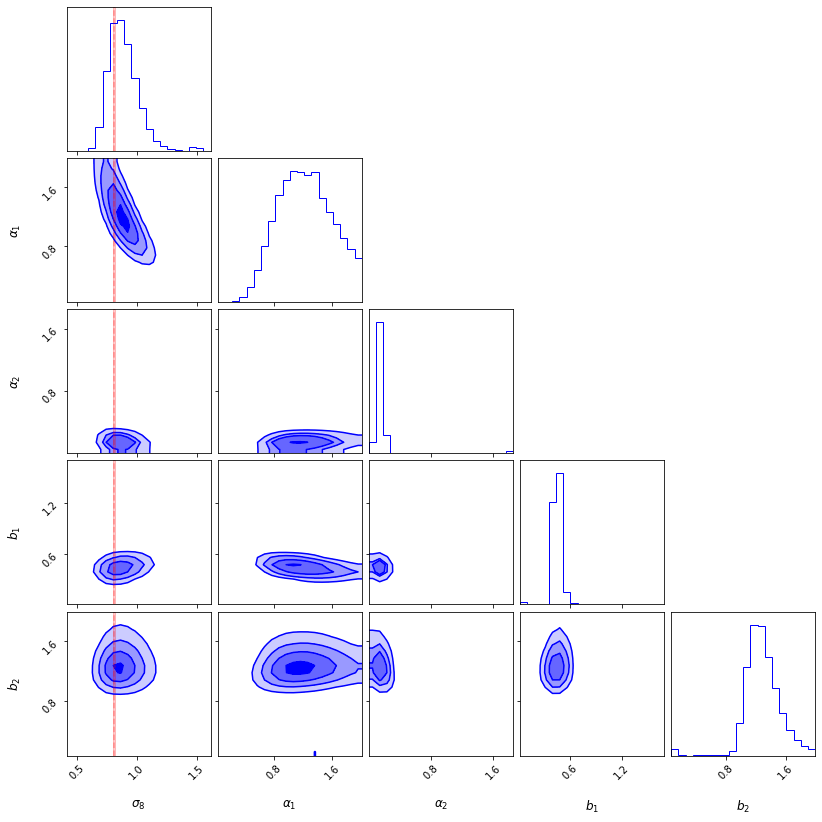}
        \caption{The posterior distributions and the contour maps for the parameters of the double lognormal model are shown. The blue contours represent confidence regions, corresponding to 1-$\sigma$ (68\%), 2-$\sigma$ (95\%), or 3-$\sigma$ (99.7\%), and 4-$\sigma$ (99.9973\%) credible intervals, respectively. 
        The red-shaded region shows the 1-$\sigma$ credible interval obtained from  \textit{Planck} 2018 TT-TE-EE model \citep{Planck:2018vyg}. The MCMC analysis was configured with 20,000 steps per walker. The first 100 steps were discarded as burn-in, and a thinning factor of 15 was applied, resulting in a total of 132,600 samples.}
        \label{fig:contour_2lognormal_Kradio}
    \end{minipage}
\end{figure*}
\begin{figure*}
 \centering
 \begin{minipage}{0.49\textwidth}
  \includegraphics[height=0.25\textheight]{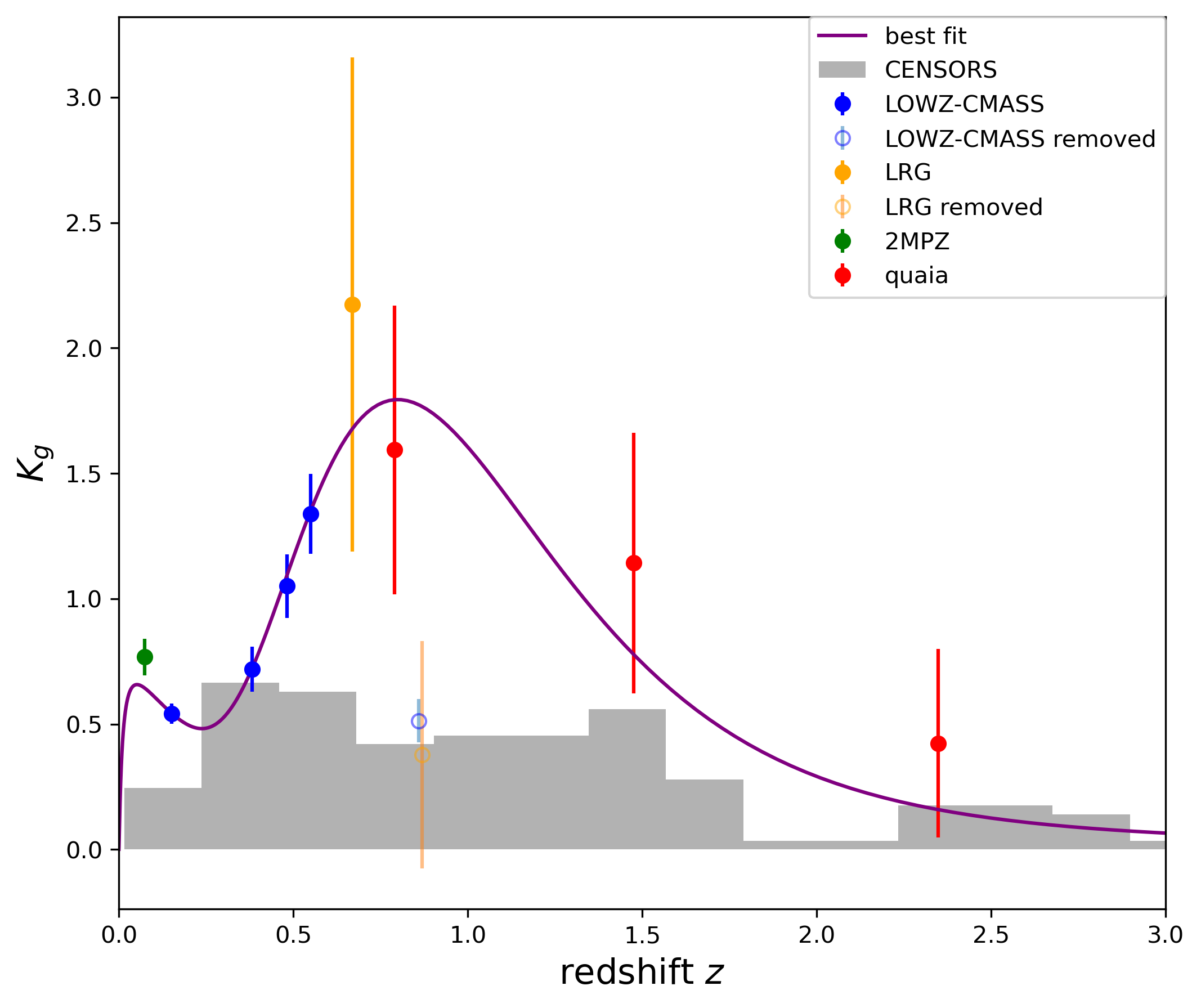}
    \end{minipage}
     \begin{minipage}{0.49\textwidth}
      \includegraphics[height=0.25\textheight]{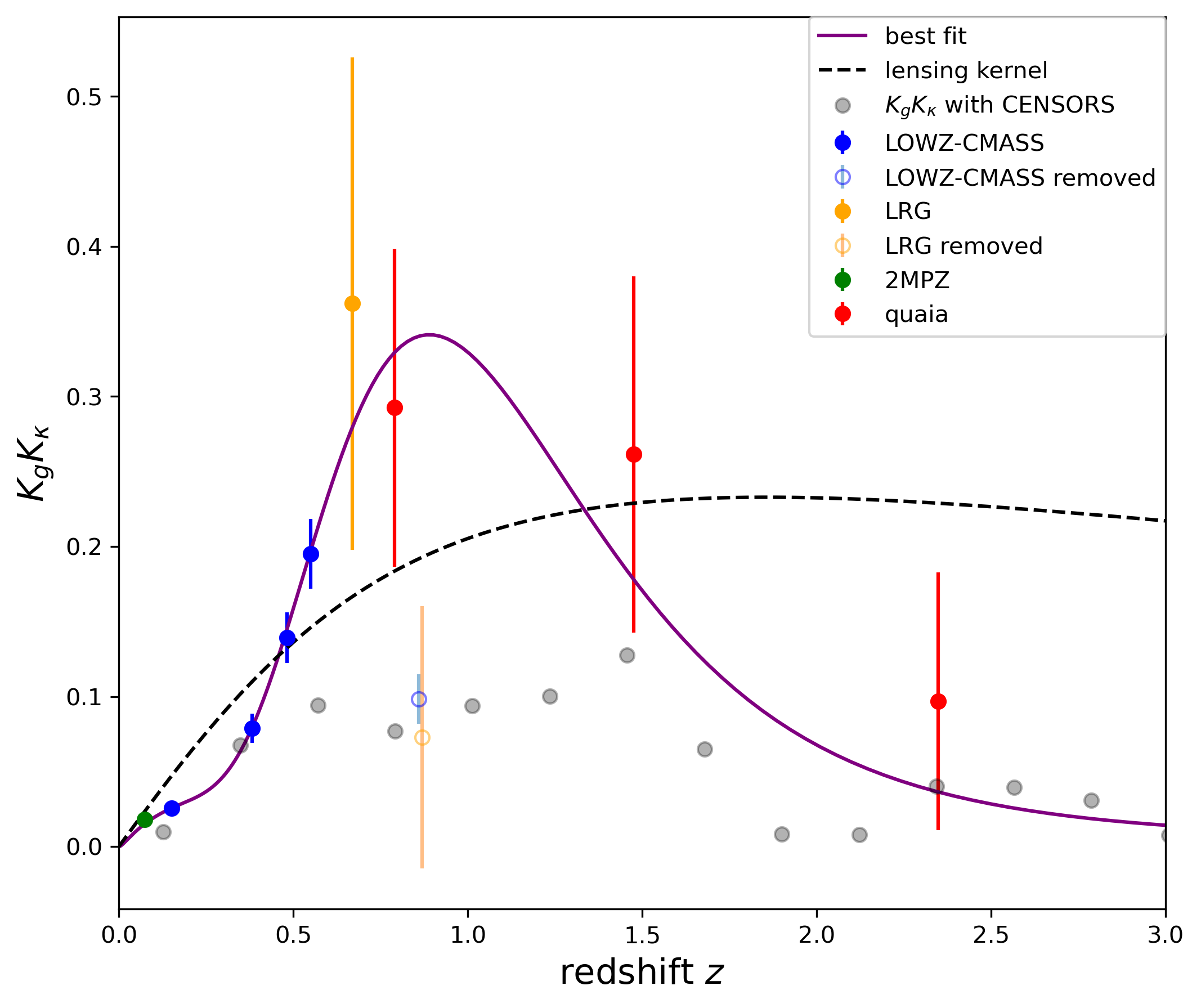}
    \end{minipage}
    \caption{Comparison of the NVSS-SUMSS kernel function (left) and the kernel of $C_{g\kappa}$ (right) with the best-fit lognormal model. The data points with filled circles show the results of reconstruction: 2MPZ (green), LOWZ-CMASS (blue), eBOSS DR16 LRGs (orange), and Gaia-unWISE (red). The two points with open circles show the reconstructed values of the kernel function determined by the removed bins: the fifth bin in LOWZ-CMASS and the second bin in eBOSS DR16 LRGs, showing the smaller amplitude. The solid thick lines show the best-fit double-lognormal function: the total (red), the primary (blue), and the secondary (green), respectively. The solid thin lines shape the double-lognormal function with the parameters $(\sigma_8,\alpha_1,\alpha_2,\beta_1,\beta_2)$ randomly selected from 20,000 realisations. As a reference, we show the CENSORS spectroscopic subsample of the NVSS galaxies by the grey histogram in the left figure and the grey-filled circles in the right figure, showing 129 objects with $>8$mJy out of 147 sources whose redshifts are obtained by either spectroscopy, $K$-$z$ estimate, $K$-$z$ limit, or $I$-$z$ estimate; cf. \citep{Brookes:2008fw}. The CENSORS catalogue is available from \url{https://vizier.cds.unistra.fr/viz-bin/VizieR-3?-source=J/MNRAS/416/1900/censors}.}
    \label{fig:reconstructed_Kg_and_KgxKk_of_NVSS-SUMSS}
\end{figure*}
\begin{figure*}
 \centering
  \includegraphics[width=12.0cm]{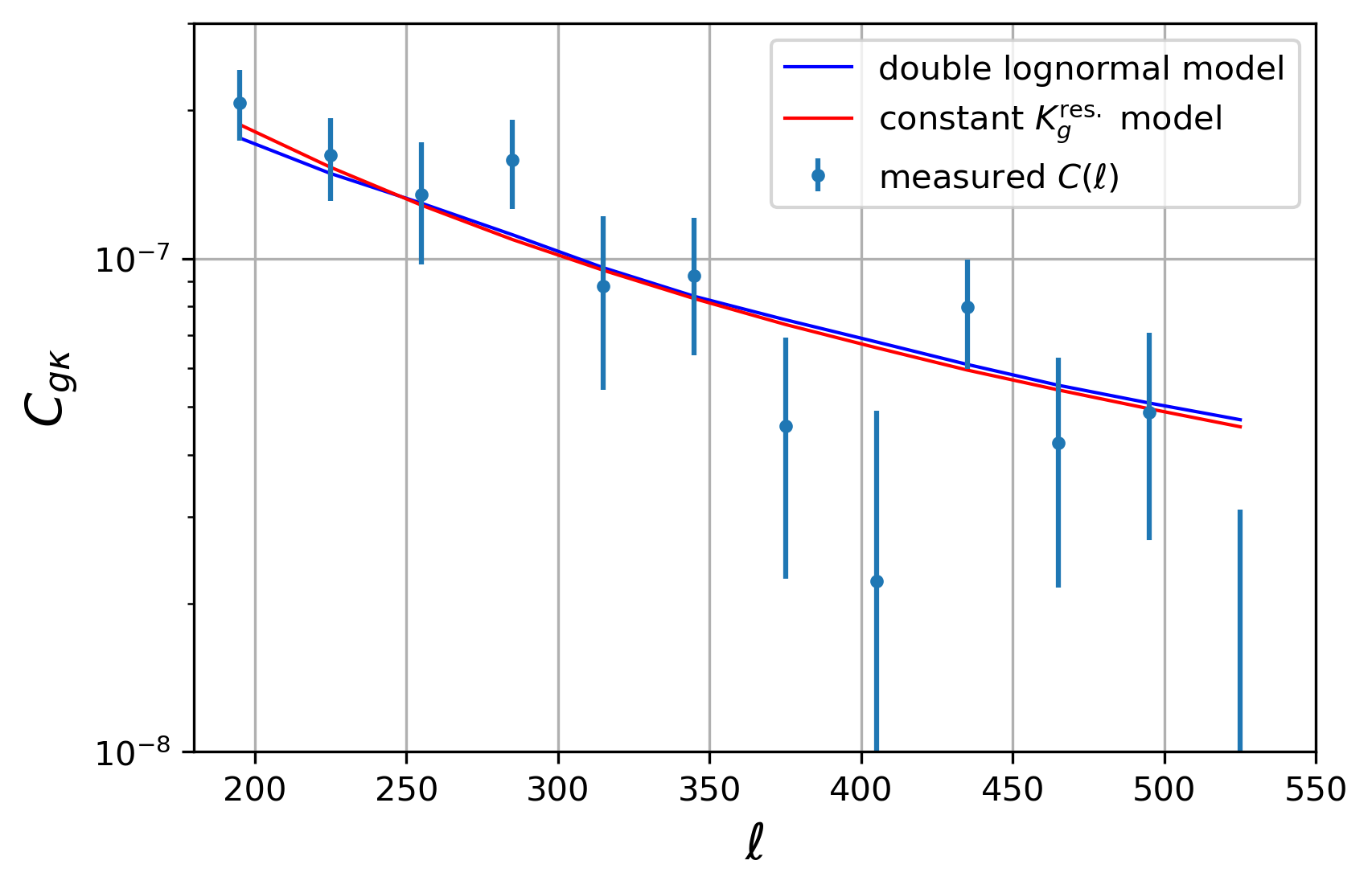}
  \caption{Comparison between the measurement of $C_{g\kappa}$ from NVSS-SUMSS $\times$ Planck PR4 and two different model of $C_{g\kappa}$. The points are the measured values using a bandwidth of $\ell_{\rm bin}=30$. The red and blue solid lines denote the predictions of $C_{g\kappa}$ from the constant $K^{\rm res.}_g$ model and the double lognormal model given the best-fit parameters, respectively. The best-fit parameters for the constant $K^{\rm res.}_g$ model and the double lognormal model are summarised in Tables.~\ref{tab:constant_Kres_bestfit} and ~\ref{tab:double_lognormal_bestfit}, respectively.}
    \label{fig:comparison_dgk_rec_vs_measurement}
\end{figure*}
\subsection{Robustness of constraint on $\sigma_8$}\label{ssec:separable_redshift}
We will further examine whether the uncertainty in the kernel function in $z > z_{\rm max.}$ affects the estimation of $\sigma_8$ given the double lognormal model. We quantify the range of redshift in which the tomographic separation is capable of given the reconstructed kernel function, calculating the signal-to-noise ratio 
\begin{equation}\label{eq:SNR_zsep}
    {\rm SNR} \equiv \sum_\ell{\left(\frac{(\hat{C}_{g\kappa}(\ell)-C_{g\kappa}(\ell,z_{\rm sep.}))^2}{\hat{\sigma}^2_{g\kappa}(\ell)} \right)} \,.
\end{equation}
Recall that $\hat{C}_{g\kappa}(\ell)$ and $\hat{\sigma}^2_{g\kappa}(\ell)$ are the measured pseudo angular cross-power spectrum and its variance, respectively. Note that $\hat{C}_{g\kappa}(\ell)$ contains the entire redshift range covered by the NVSS-SUMSS sample.
On the other hand, $C_{g\kappa}(\ell,z_{\rm sep.})$ is calculated as
\begin{equation}
    C_{g\kappa}(\ell,z_{\rm sep.}) \equiv \int^{z_{\rm sep.}}_{0}{\rm d}z\frac{K^{{\rm model}}_g(z)K_\kappa(z)}{\chi^2(z)c/H(z)}P\left(\frac{\ell+1/2}{\chi(z)},\,z\right)\,.
\end{equation}
$C_{g\kappa}(\ell,z_{\rm sep.})$ describes the truncated $C_{g\kappa}$ up to $z = z_{\rm sep.}$. Subtracting it from the measured $C_{g\kappa}(\ell)$ can then give an indication of the residual contribution to CMB lensing sourced by the NVSS-SUMSS galaxies at $z > z_{\rm sep.}$. In this sense, the SNR in Eq.~\eqref{eq:SNR_zsep} explicitly assesses whether the residual signal is below or above the noise.

Given the SNR we can determine $z_{\rm sep.}$ by solving the above formula for $z_{\rm sep.}$. Since $\smash{K^{{\rm model}}_g(z)}$ depends on the parameters $(\sigma_8,\alpha_1,\alpha_2,\beta_1,\beta_2)$,  $z_{\rm sep.}$ therefore has this same dependence.
Note that $C_{g\kappa}(\ell,z_{\rm sep.})$ is monotonic as $z_{\rm sep.}$ increases, therefore there is a unique solution for $z_{\rm sep.}$ for each SNR.
Before calculating $z_{\rm sep.}$ from Eq.~\eqref{eq:SNR_zsep}, we compute the cumulant $c(\ell,z)\equiv C_{g\kappa}(\ell,z)/C_{g\kappa}(\ell,z_{\rm sep.\ max.})$, which quantitatively shows the percentage of each tomographic bins as a function of redshift. In Fig.~\ref{fig:cumulant_cgk}, we show $c(\ell,z)$ for a selection of samples from the 132,600 MCMC realisations at $\ell=225$. We set $z_{\rm sep.\ max.}=10$ above which the cumulant converges well to unity. Note that we confirmed that the other multipoles deviate little from each other in showing the main feature.
We find that the cumulant is less than unity, converging around $z \approx 4$. This indicates that the NVSS-SUMSS radio samples may provide the deepest tomographic dissection of the CMB lensing signals in redshift, extending the previous results of \cite{Peacock:2018xlz,2021JCAP...12..028K,2023JCAP...11..043A, 2024JCAP...06..012P} to higher redshifts.

We then assess how much the higher redshift information can be separable with the PR4 $\times$ radio angular cross-power spectrum measurement errors. Fig.~\ref{fig:separable_redshifts_given_noise_from_cgk_PR4_NVSS-SUMSS} shows the derived $z_{\rm sep.}$ as a function of signal-to-noise ratio. We find that $z_{\rm sep.} \approx 2$ is the redshift at which one can separate the tomographic slices from the measured $\hat{C}_{g\kappa}$ obtained by the NVSS-SUMSS $\times$ {\it Planck} PR4 given ${\rm SNR} = 1$. Regarding that we made the tomographic reconstruction of the NVSS-SUMSS kernel function up to $z \approx 2.3$, this means that there is no evidence for residual signals explaining the measured $C_{g\kappa}(\ell)$ given the present dataset. Recalling that the residual component systematically makes $\sigma_8$ smaller given Eq.~\eqref{eq:alpha_sigma8_estimator}, we conclude that such a bias from the residual component does not affect the measurement of $\sigma_8$ given the sensitivity of the \textit{Planck} PR4 CMB lensing convergence.

%% file: App.tex
\section{Estimator of auto-power spectrum tomography with reconstructed kernel function}\label{app:auto-power_spectrum_tomography_with_rec_Ku}
We supply here the formulae of the angular auto-power spectrum of a lens-mass sample, given that the clustering-redshift analysis reconstructs a kernel function of the sample. We denote the lensing sample with the subscript $u$ following the main text. The angular auto-power spectrum of the lens-mass sample is given as
\begin{equation}\label{eq:cluu_limber}
    C_{uu}(\ell) = \int^{\infty}_0{{\rm d}z} \frac{K^2_u(z)}{\chi^2c/H} P_m\left(\frac{\ell+1/2}{\chi},\,z\right)\,.
\end{equation}
Following the logic similar to deriving $C_{u\kappa}$ in Sec.~\ref{ssec:CMB lensing_tomography_with_rec_Ku}, we obtain
\begin{align}\label{eq:cl_uu_dec}
    C_{uu}(\ell) \sim\  & \Sigma_{uu}(\ell) + S_{uu}(\ell)\,,\\ \nonumber
    \Sigma_{uu}(\ell) \equiv& \sum^{N_{\rm rec.}}_{i=1}\frac{K^{\rm rec. 2}_u(z_i)\delta z_i}{\chi^2(z_i)c/H(z_i)}P_m\left(\frac{l+1/2}{\chi(z_i)},z_i\right)\,,\\ \nonumber
    S_{uu}(\ell) \equiv&  \int^{z_*}_{{z_{\rm max}}}{ {\rm d}z\frac{{K^{\rm res. 2}_u}(z)}{\chi^2(z)c/H(z)} P_m\left(\frac{l+1/2}{\chi(z)},z\right)}\,.
\end{align}
Note that the notation follows  Sec.~\ref{ssec:CMB lensing_tomography_with_rec_Ku}. We find that $\Sigma_{uu}$ is proportional to $\sigma_{8,{\rm true}}\sigma_{8,{\rm fid.}}$, substituting $K^{\rm rec.}_u \propto \sigma_{8,{\rm true}}/\sigma_{8,{\rm fid.}}$ and $P_m \propto \smash{\sigma^2_{8,{\rm fid.}}}$. On the other hand, the left-hand side $C_{uu}(\ell)$ is proportional to $\smash{\sigma^2_{8,{\rm true}}}$, which can be measured using two-point statistics. Then we define
\begin{align}\label{eq:beta_sigma8_estimator}
    \beta \equiv \frac{C_{uu}}{\Sigma_{uu} + S_{uu}}\,,
\end{align}
estimating $\sigma_{8,{\rm true}}$ independently from $\alpha$ in Eq.~\eqref{eq:alpha_sigma8_estimator}. Provided $S_{uu}$ is negligibly small compared to $\Sigma_{uu}$, we obtain $\beta = C_{uu}/\Sigma_{uu}$ and $\sigma_{8,{\rm true}} = \beta\sigma_{8,{\rm fid.}}$. Since $\beta$ consists only of the measured two-point statistics, a similar conclusion applies to $\alpha$ provided that $S_{uu}$ is negligible compared to $\Sigma_{uu}$. One can apply these formulae given a measurement of the auto-power spectrum of the lensing sample, though this is not the case with the NVSS-SUMSS dataset.%
\section{Validation of linear bias model in configuration space analysis}\label{app:validation_of_linear_bias_model}
We will confirm that the galaxy biases of the reference samples are independent of scale over an angular range determined by a certain choice of scale cuts. We consider the scale cut in the range $\theta \in [0.3^\circ, 1^\circ]$, the same as the definition in the main analysis. 

\begin{figure*}
 \centering
  \includegraphics[width=12.0cm]{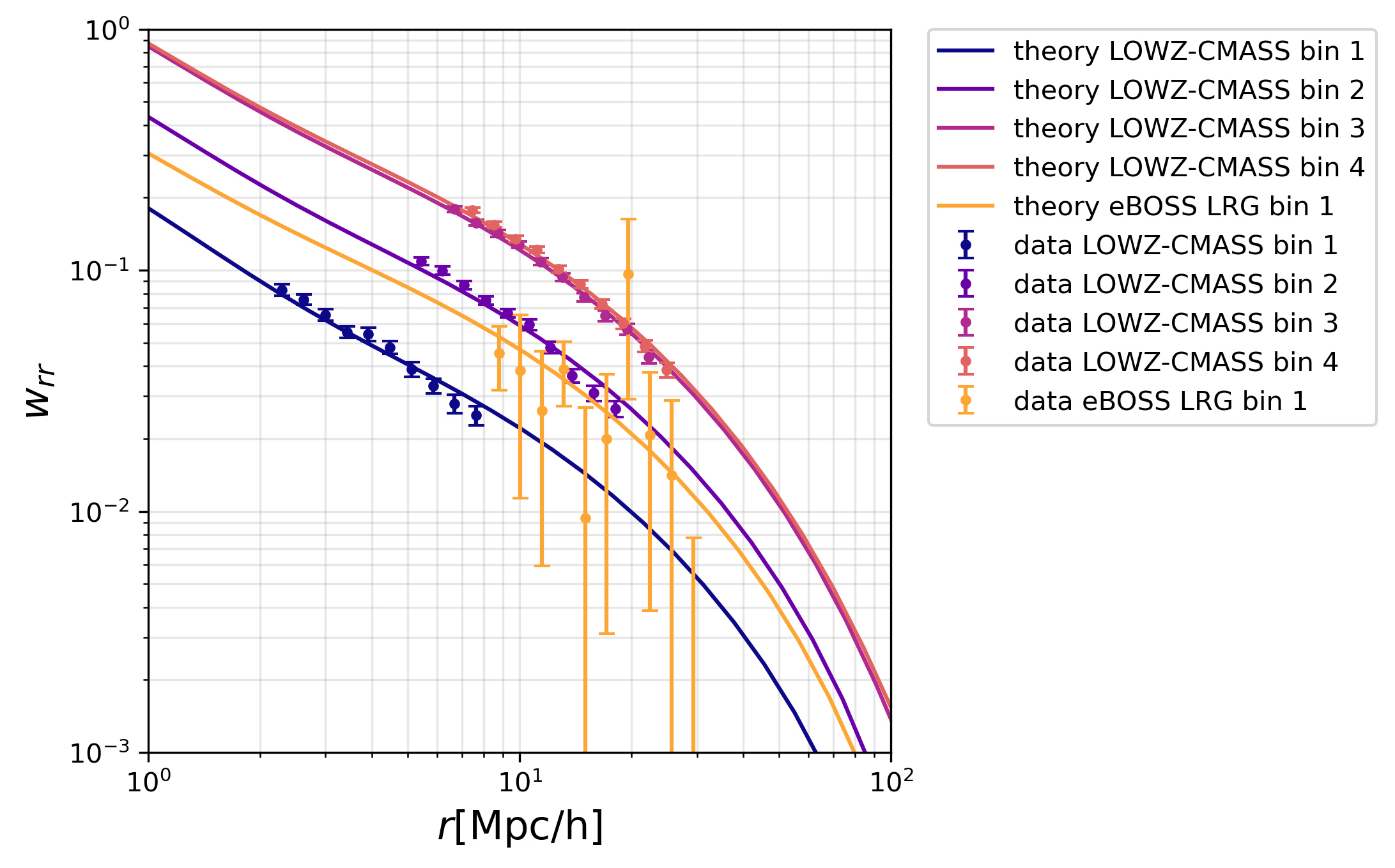}
  \caption{The auto-correlation functions of the reference galaxies. The solid lines are theoretical predictions with estimated bias $b_r$ of each bin. The data points span between $0.3 ^\circ < \theta < 1^\circ $ where the measurements are carried out. The colours of the solid lines and the data points are consistent for a given sample.}
    \label{fig:corr_CMASS5_LRG2}
\end{figure*}

\begin{figure*}
 \centering
  \includegraphics[width=12.0cm]{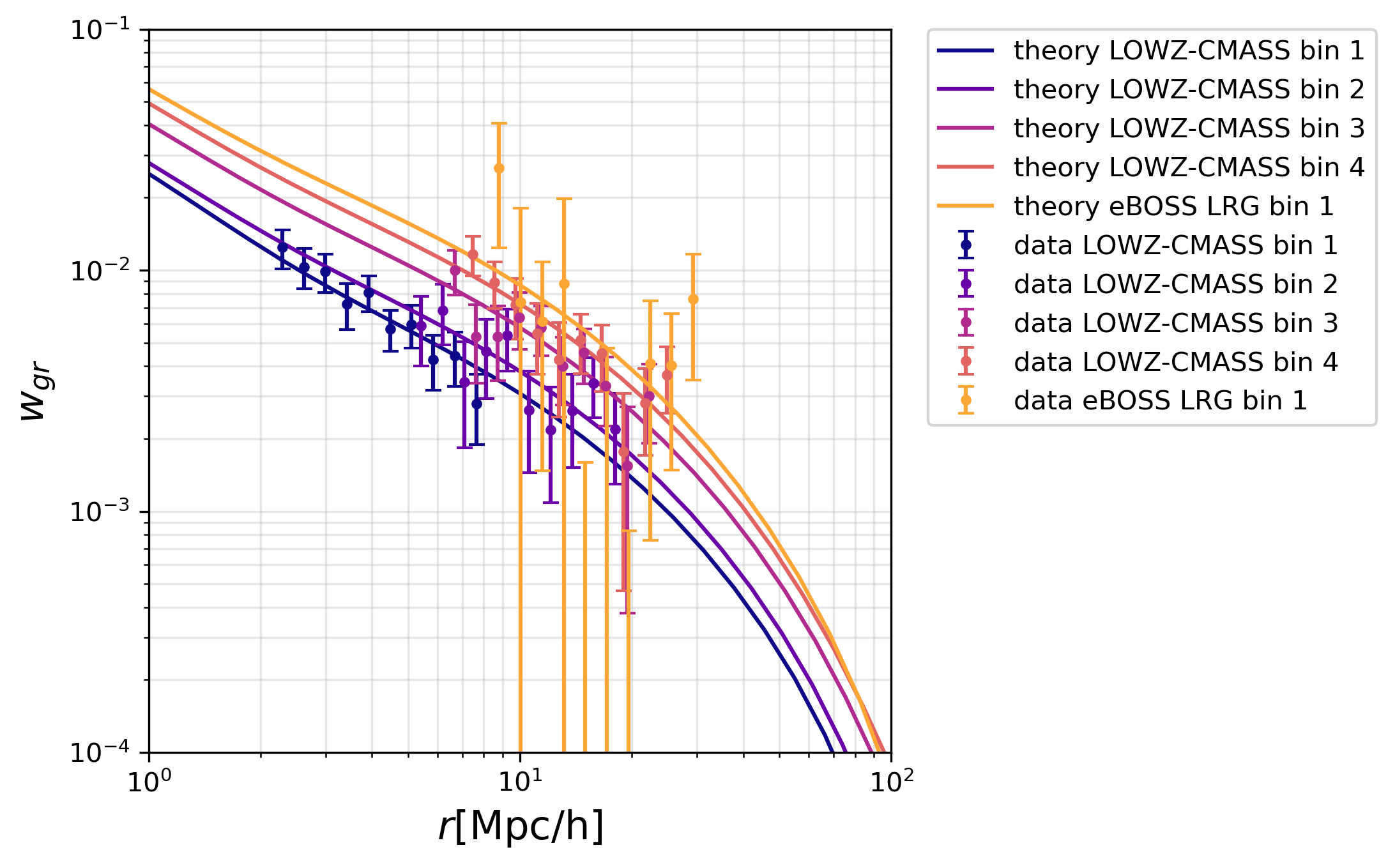}
  \caption{The cross-correlation functions between the NVSS-SUMSS galaxies and the reference galaxies. The solid lines are theoretical predictions with estimated $b_r K_g$ of each bin. The data points span between $0.3\ ^\circ < \theta < 1^\circ $ where the measurements are carried out. The colours of the solid lines and the data points are consistent for a given sample, as well as with the colours for the auto-correlation functions in Fig.~\ref{fig:corr_CMASS5_LRG2}.}
    \label{fig:xcorr_NVSS_SUMSS_CMASS5_LRG2}
\end{figure*}

\begin{figure*}
 \centering
  \includegraphics[width=12.0cm]{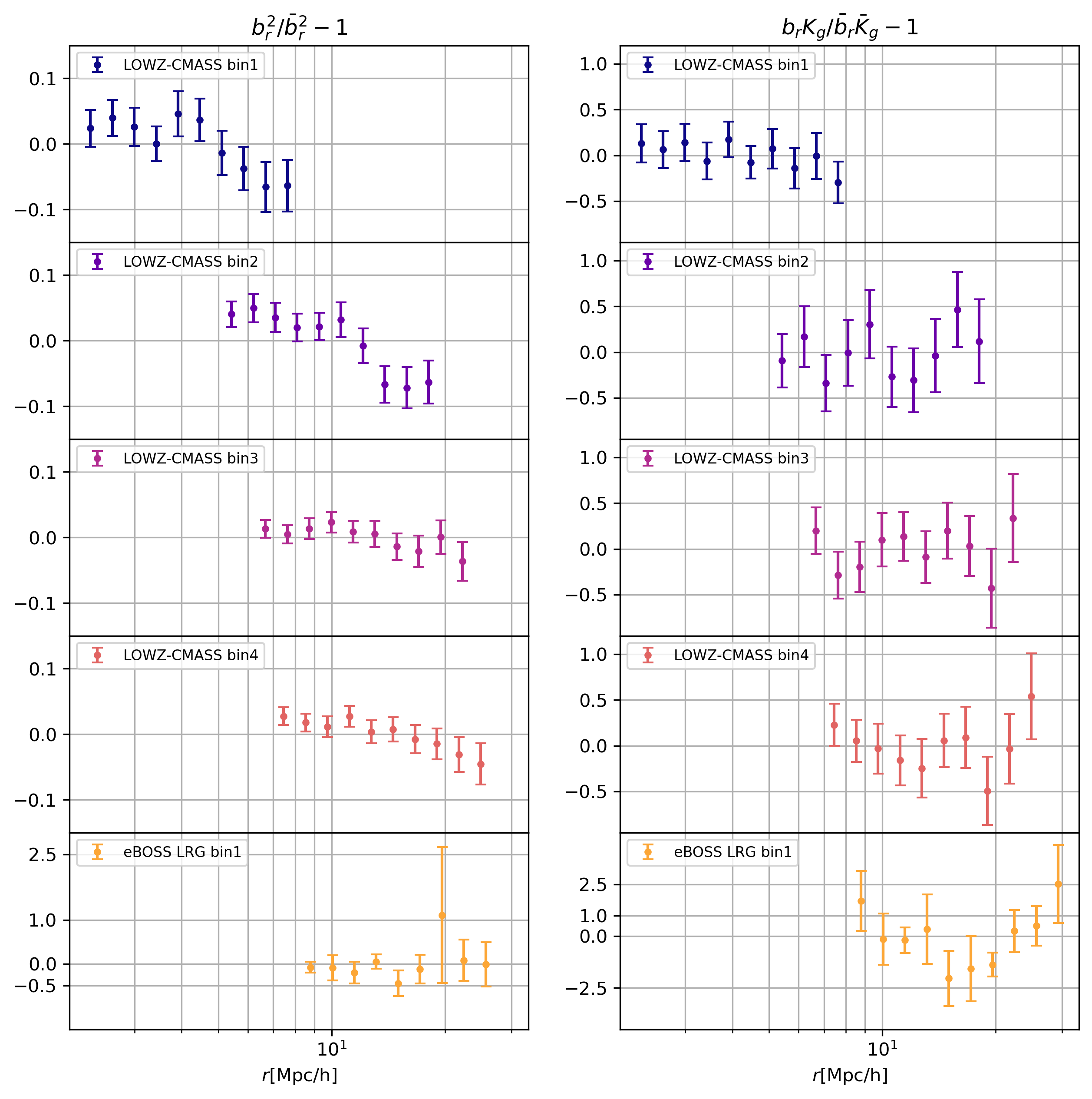}
  \caption{The scale dependence of the measured linear biases (left column) and the radio kernel functions (right column), plotting the normalised deviation of the measured amplitude from the mean values as shown in the titles of the columns. 
  The last row for the left and right columns is scaled differently in the vertical axes. The colours of the data points are the same as those for the auto-correlation functions in Figs~\ref{fig:corr_CMASS5_LRG2} and ~\ref{fig:xcorr_NVSS_SUMSS_CMASS5_LRG2}.}
    \label{fig:check_scale_dependence_configspace}
\end{figure*}

Figs~\ref{fig:corr_CMASS5_LRG2} and ~\ref{fig:xcorr_NVSS_SUMSS_CMASS5_LRG2} compare the theoretical predictions for the auto-correlation functions and the cross-correlation functions with the measurements, assuming the fiducial cosmological model plus  the HALOFIT model of the nonlinear matter-power spectrum. Figs~\ref{fig:corr_CMASS5_LRG2} and ~\ref{fig:xcorr_NVSS_SUMSS_CMASS5_LRG2} display a good agreement between the theory and the data, though it is not clear whether the amplitude can be approximated in a scale-independent way. Hence we carry out the analysis for the deviation from the mean i.e. $b_r/\bar{b}_r-1$ and $b_rK_g/\bar{b}_r\bar{K_g}-1$, as shown in Fig.~\ref{fig:check_scale_dependence_configspace}.

Fig.~\ref{fig:check_scale_dependence_configspace} provides a test of the scale-dependence of the amplitudes. The measured values of $b_r$ and $b_rK_g$ show a flat dependence on scale within acceptable stochastic fluctuations around the mean, though the last row for the LRGs is noisy. We find that the LOWZ-CMASS bins suggest 10\% precision in $b_r$ and 50\% precision in $b_r K_g$, respectively. The LRGs bin is the most noisy, though the data points do not show distinguishable evidence for scale dependence. In short, this establishes the measured quantities that satisfy our assumptions made in reconstructing the radio kernel function.

\section{Fitting of cross-noise model in harmonic space analysis}\label{app:cross_noise_in_detail}
We supply the posterior distributions for the parameters of the cross-noise model described in Sec.~\ref{sssec:est_harmo-sp_analysis}, fitting the measured angular cross-power spectra between the NVSS-SUMSS radio galaxies and the photometric tracers i.e. 2MPZ and Gaia-unWISE QSOs. We apply a multipole scale cut for 2MPZ as a restriction to $20 \leq \ell \leq 250$, with the cut for the Gaia-unWISE QSO sample being $180 \leq \ell \leq 540$ and $s=0.8$, both of which are equivalent in defining the reference samples to those used in the main analysis.

\begin{figure}
 \centering
  \includegraphics[width=7.5cm]{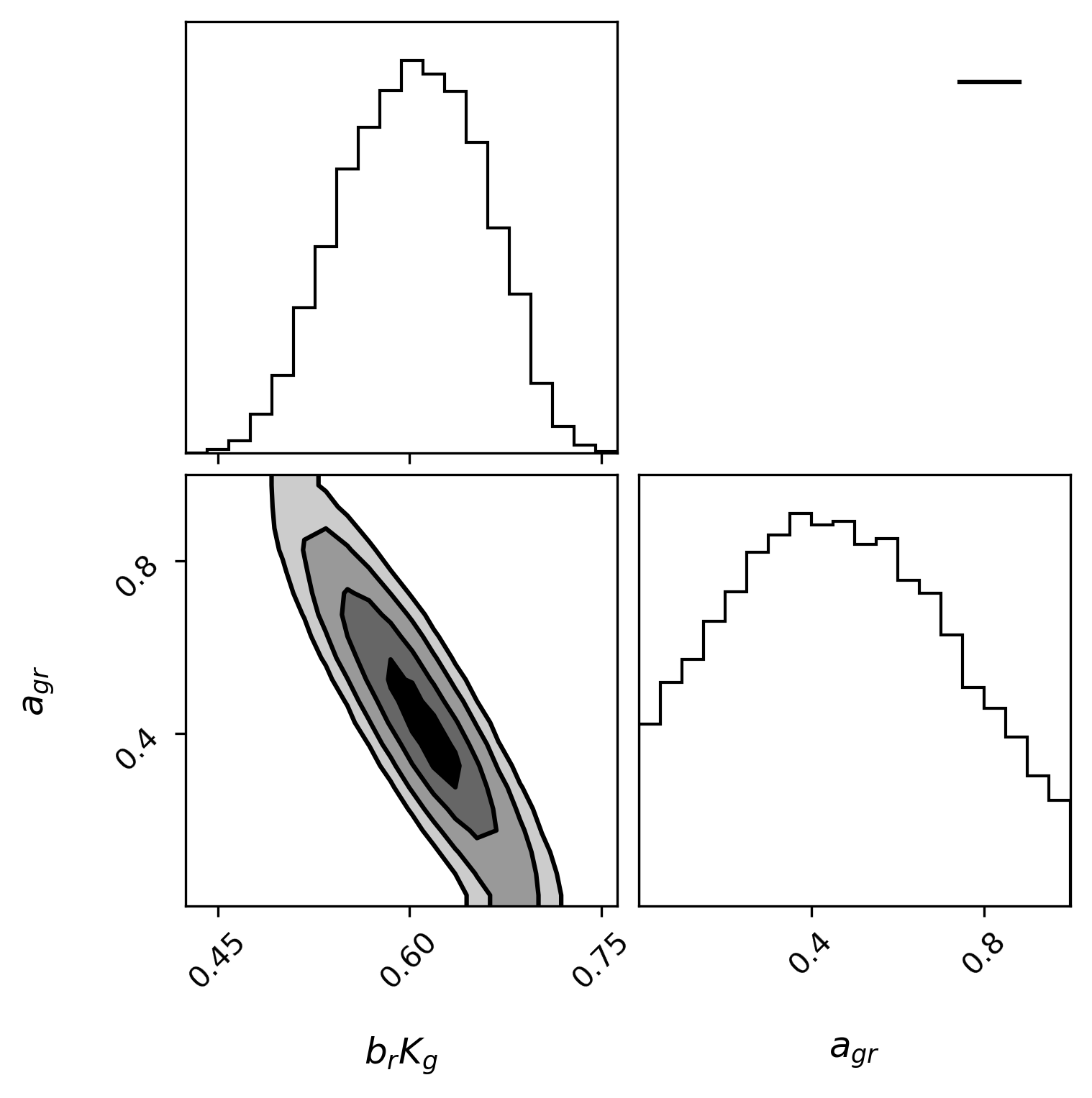}
  \caption{The posterior distribution of the fitted parameters in the cross-noise model for the angular cross-power spectrum between the NVSS-SUMSS galaxies and the 2MPZ galaxies.}
    \label{fig:cross-noise_modelfit_2MPZ_lmax250}
\end{figure}
\begin{figure}
 \centering
  \includegraphics[width=7.5cm]{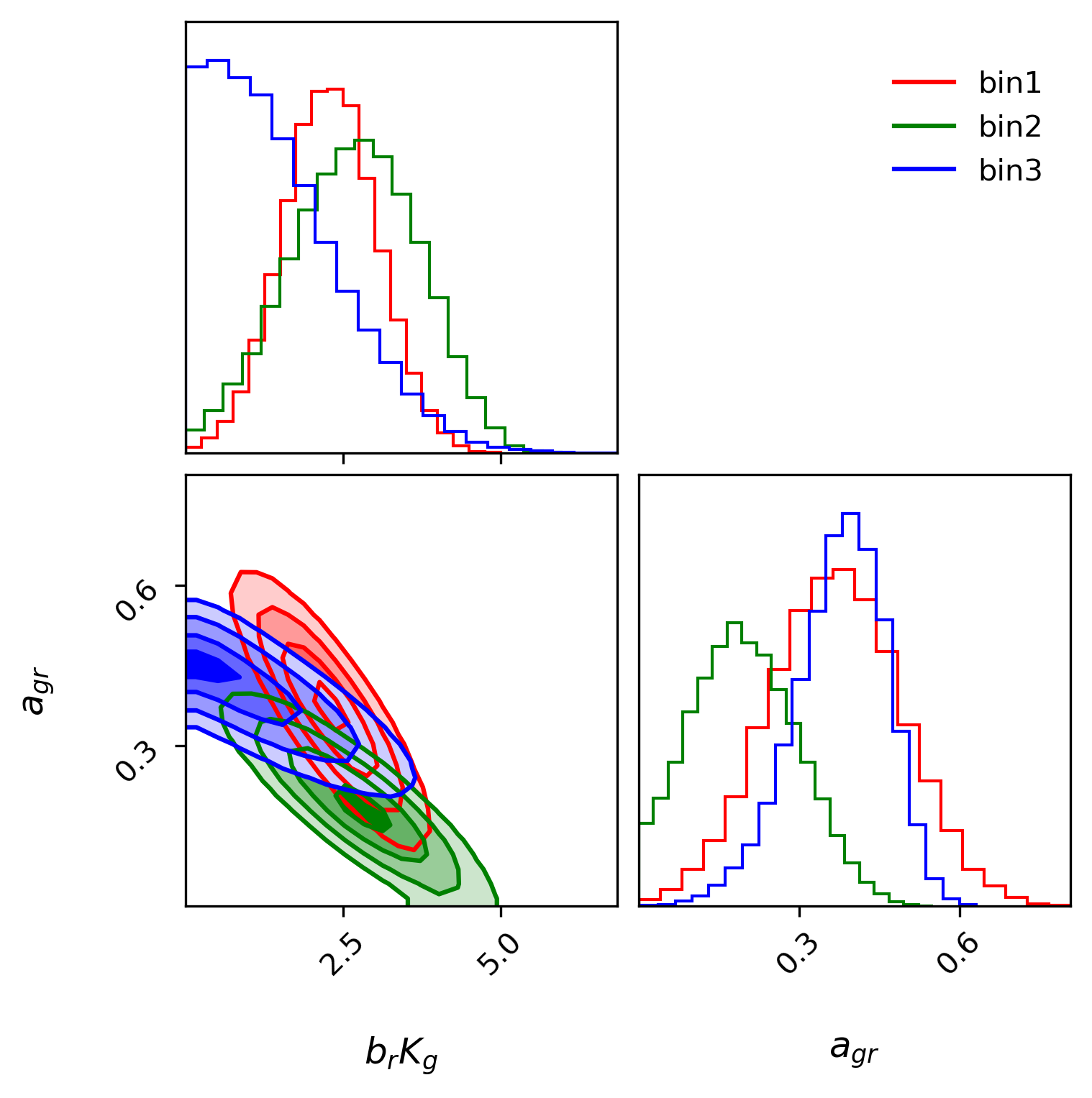}
  \caption{The posterior distribution of the fitted parameters in the cross-noise model for the angular cross-power spectrum between the NVSS-SUMSS galaxies and the Gaia-unWISE galaxies. The red, green, and blue contours shows the error regions for the first, second, and third redshift bins of the Gaia-unWISE QSO samples, respectively.}
    \label{fig:cross-noise_modelfit_Gaia-unWISE_QSOs}
\end{figure}

\begin{figure}
 \centering
  \includegraphics[width=7.0cm]{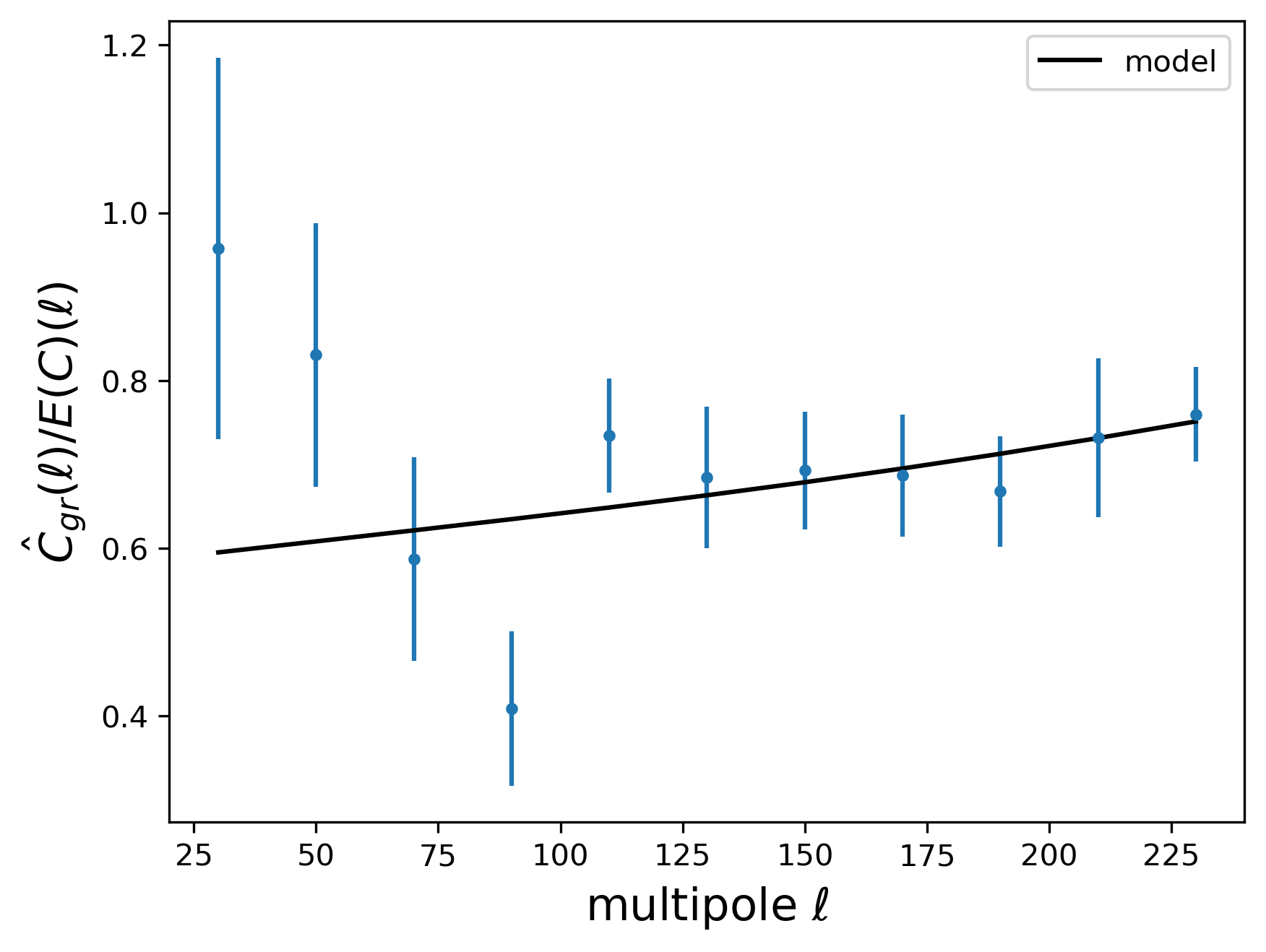}
  \caption{Multipole dependence of the linear bias of 2MPZ galaxies ${\hat{C}}_{ur}(\ell)/E(C)(\ell)$ given $20 \leq \ell \leq 250$.}
    \label{fig:multipole_dependence_b2MPZ}
\end{figure}

\begin{figure}
 \centering
  \includegraphics[width=7.0cm]{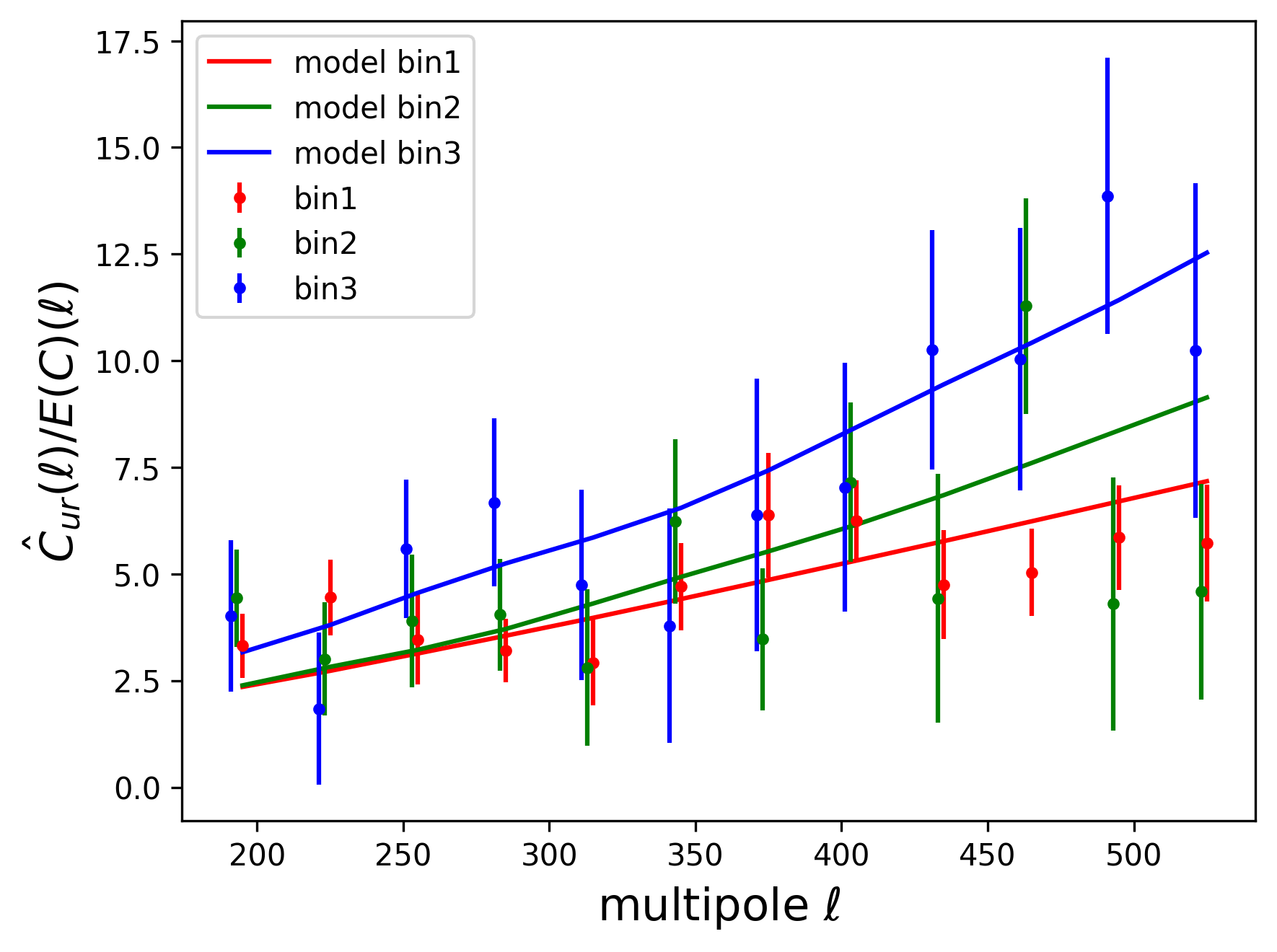}
  \caption{Multipole dependence of ${\hat{C}}_{ur}(\ell)/E(C)(\ell)$ given $s=0.8$ and $180 \leq \ell \leq 540$.}
    \label{fig:multipole_dependence_bqsoxKradio}
\end{figure}

Fig.~\ref{fig:cross-noise_modelfit_2MPZ_lmax250} shows the error contour of the parameters of the cross-noise model for the NVSS-SUMSS and the 2MPZ angular cross-power spectrum.  Fig.~\ref{fig:cross-noise_modelfit_Gaia-unWISE_QSOs}, on the other hand, shows the error contours of the parameters of the cross-noise model for the NVSS-SUMSS and the Gaia-unWISE angular cross-power spectrum, separated into the three redshift bins of the QSO samples. One can see that the cross-noise amplitude $a_{gr}$ is non-zero high significance, tracing the cross-noise component from the data. The best-fit values of the parameters of the cross-noise model are shown in Table~\ref{tab:cross_noise}. Figs~\ref{fig:multipole_dependence_b2MPZ} and ~\ref{fig:multipole_dependence_bqsoxKradio} compare the best-fit values and the measured angular cross-power spectrum divided by $E(C)$ between 2MPZ and NVSS-SUMSS and between Gaia-unWISE QSOs and NVSS-SUMSS, respectively. One can see that the prediction of  Eq.~\eqref{eq:cross_noise_model} traces the scale dependence of the data rather accurately, which helps subtract the cross-noise component by fitting $a_{gr}$.


\section{Effects of various sample cut for the Gaia-unWISE sample}\label{app:sample_cut_quaia}

We show our results are affected by exactly how we define a Gaia-unWISE QSO sample with a given selection function $s$ to derive the reconstructed radio kernel function.

\begin{figure*}
\vspace{2.0cm}
 \centering
  \includegraphics[width=10.0cm]{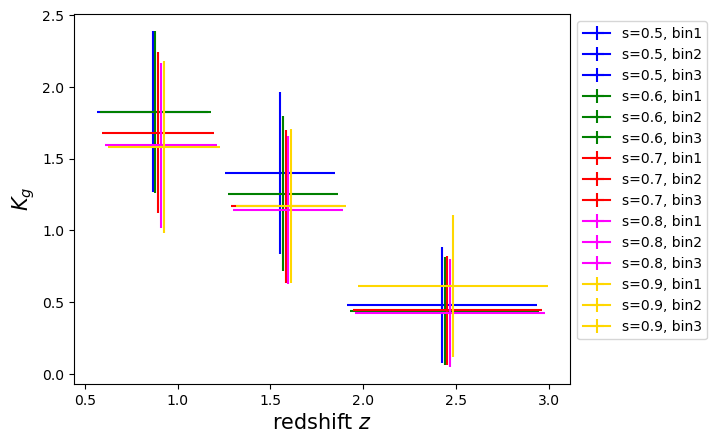}
  \caption{Reconstructed kernel functions with different values of the selection function.}
    \label{fig:rec_kernels_sel_thrshld_variations}
\end{figure*}

Firstly, we examine how the threshold value for the selection function $s$ of the Gaia un-WISE QSO samples affects the amplitudes of the reconstructed kernel function. Fig.~\ref{fig:rec_kernels_sel_thrshld_variations}
shows that there is no strong dependence on the selection threshold for the analysis: even though some systematic decrease of the amplitudes in the two smaller redshift bins is observed, all the points are consistent within the error bars. We choose a selection threshold $s=0.8$ as in the main analysis, which can be conservative in ensuring the purity of a QSO catalogue. From this, we conclude that the various definitions of data samples have little impact on our main results. Note that we fix the selection threshold $s=0.8$ for all the variations.

\begin{figure*}
 \centering
  \includegraphics[width=10.0cm]{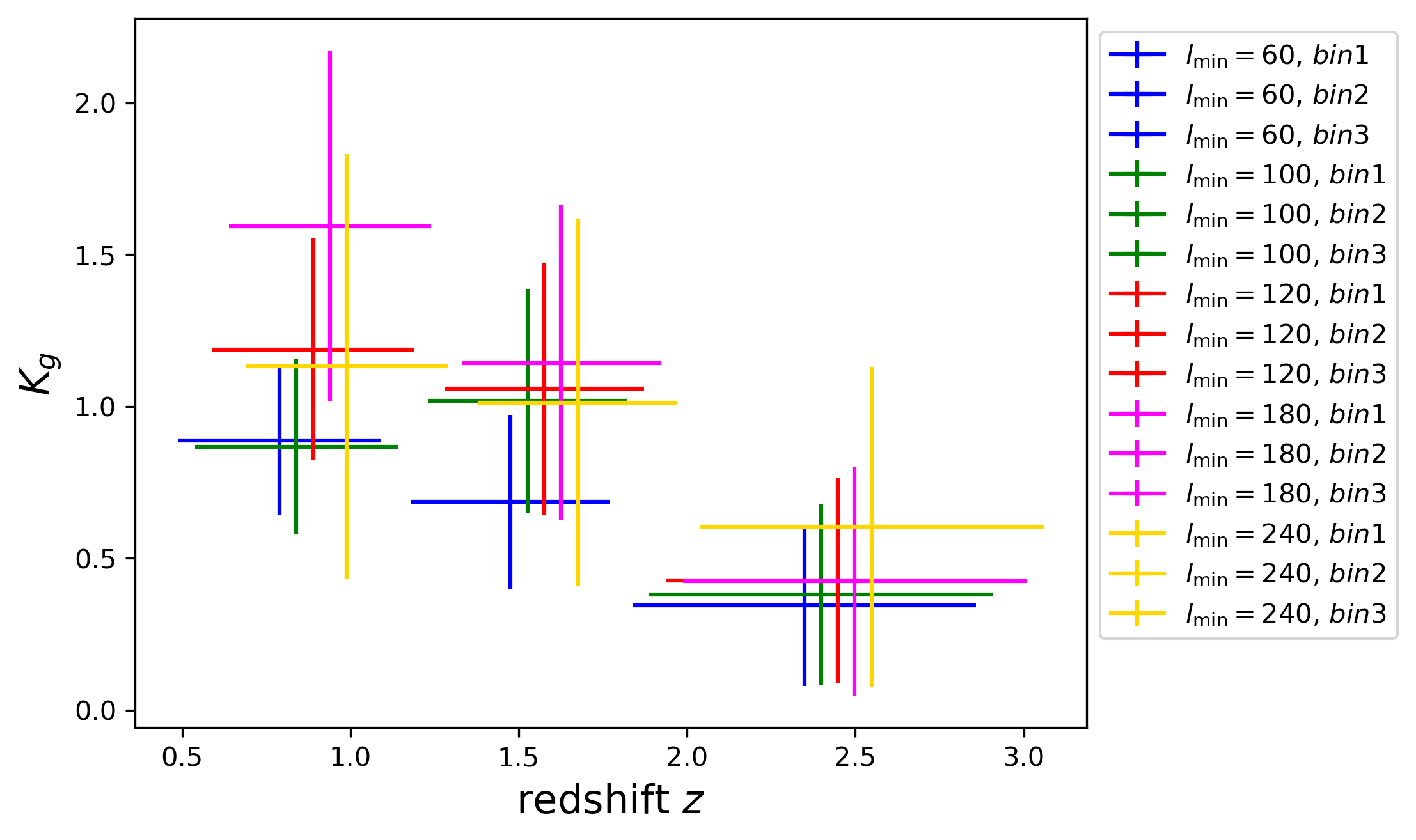}
  \caption{Reconstructed kernel functions with different values of the lowest multipole cut.}
    \label{fig:rec_kernels_lmin_variations}
\end{figure*}

Secondly, we investigate whether or not the scale cut of the minimum multipole $\ell_{\rm min.}$ has any affect on the amplitude of the reconstructed kernel function.
Fig.~\ref{fig:rec_kernels_lmin_variations} shows there is a systematic decrease in the amplitudes of the reconstructed kernel functions for smaller $\ell_{\rm min}$, while $\ell_{\rm min}=100, 180, 240$ show a relatively stable amplitude. The error bars enlarge for larger $\ell_{\rm min}$ as the number of bins decreases. Following \cite{2023JCAP...11..043A}, we presume that the leakage from the dust emission had some effect on the QSO maps. To avoid the leakage while retaining a reasonable size of the error bar, we ended up choosing $\ell_{\rm min.}=180$ for all analyses in the main text. 

\begin{figure*}
 \centering
  \includegraphics[width=10.0cm]{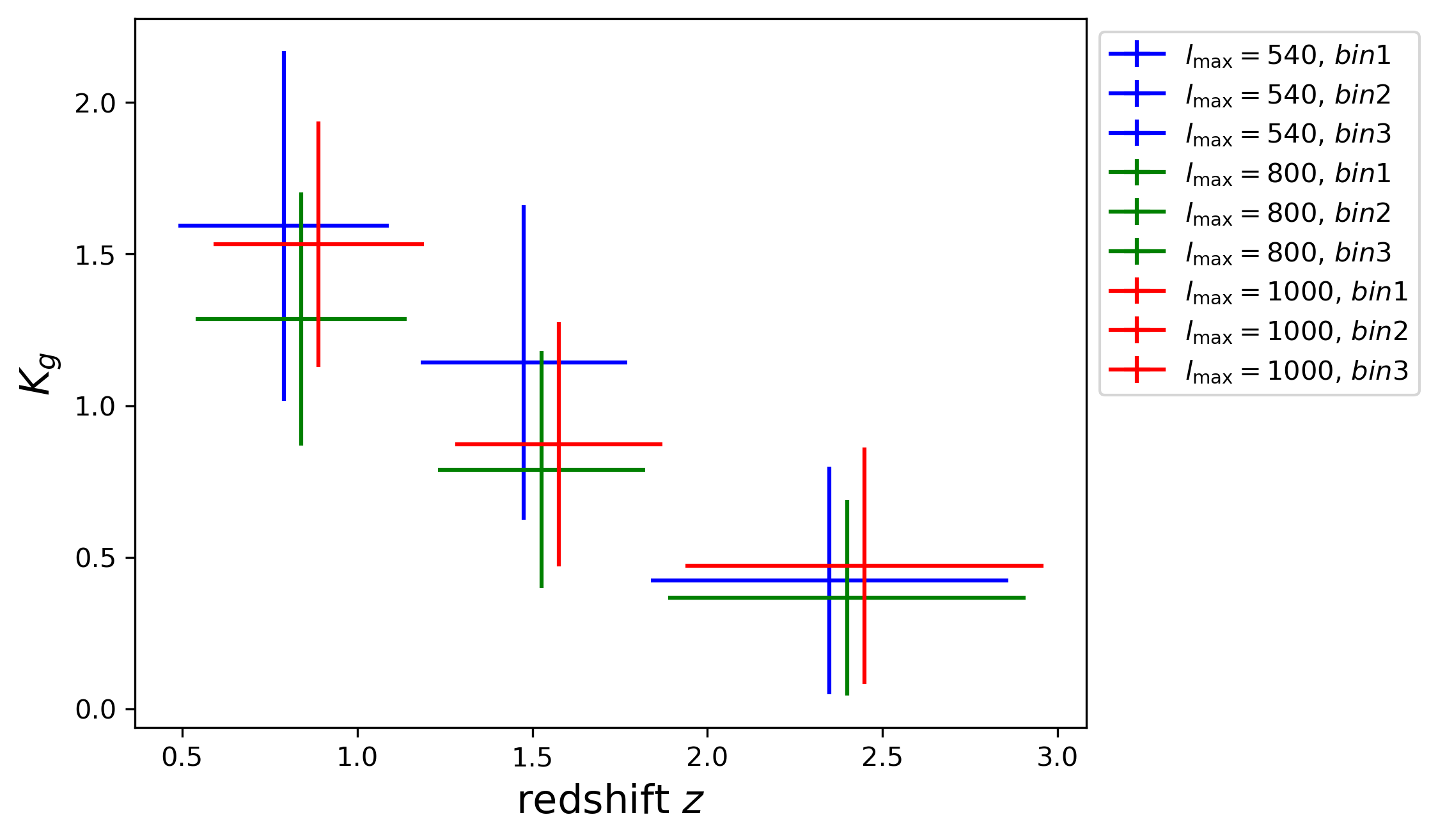}
  \caption{Reconstructed kernel functions with different values of the highest multipole cut.}
    \label{fig:rec_kernels_lmax_variations}
\end{figure*}

Finally, we similarly examine the impact of the choice of maximum multipole $\ell_{\rm max.}$ on the amplitude of the reconstructed kernel function. Fig.~\ref{fig:rec_kernels_lmax_variations} shows there is little sensitivity to this parameter in practice. Note that raising $\ell_{\rm max.}$ causes the 2-halo approximation to become less valid. The value of $\ell_{\rm max.}$ for our main analysis is therefore set so as to mitigate the inaccuracy of the dark-matter halo modelling on smaller scales.

%